\documentclass[amsmath,amssymb,amsfonts,aps,prb,twocolumn,showpacs,superscriptaddress]{revtex4-1}

\usepackage{graphicx}
\usepackage{dcolumn}
\usepackage{bm}
\usepackage{color}
\usepackage[normalem]{ulem}

\begin{document}

\newcommand{\bra}[1]{\left\langle#1\right|}
\newcommand{\ket}[1]{\left|#1\right\rangle}
\newcommand{\bracket}[2]{\big\langle#1 \bigm| #2\big\rangle}

\newcommand{\Tr}{{\rm Tr}}
\renewcommand{\Im}{{\rm Im}}
\renewcommand{\Re}{{\rm Re}}
\newcommand{\sgn}{{\rm sgn}}

\newcommand{\LL}{{\rm L}}
\newcommand{\RR}{{\rm R}}

\newcommand{\xc}{{\rm xc}}
\newcommand{\Hxc}{{\rm Hxc}}
\newcommand{\CIM}{{\rm CIM}}
\newcommand{\SSM}{{\rm SSM}}
\newcommand{\inter}{{\rm inter}}
\newcommand{\Hund}{{\rm Hund}}
\newcommand{\ave}{{\rm ave}}
\newcommand{\ske}{{\rm skew}}

\newcommand{\su}{\uparrow}
\newcommand{\sd}{\downarrow}

\newcommand{\tip}{{\rm tip}}

\title{Exchange-correlation potentials for multi-orbital quantum dots subject to generic density-density interactions and Hund's rule coupling}

\author{Nahual Sobrino}
\affiliation{Donostia International Physics Center (DIPC), Paseo Manuel de
  Lardizabal 4, E-20018 San Sebasti\'{a}n, Spain}
\affiliation{Nano-Bio Spectroscopy Group and European Theoretical Spectroscopy
Facility (ETSF), Dpto. de F\'{i}sica de Materiales,
Universidad del Pa\'{i}s Vasco UPV/EHU, Av. Tolosa 72,
E-20018 San Sebasti\'{a}n, Spain}

\author{Stefan Kurth}
\affiliation{Nano-Bio Spectroscopy Group and European Theoretical Spectroscopy
Facility (ETSF), Dpto. de F\'{i}sica de Materiales,
Universidad del Pa\'{i}s Vasco UPV/EHU, Av. Tolosa 72,
E-20018 San Sebasti\'{a}n, Spain}
\affiliation{IKERBASQUE, Basque Foundation for Science, Maria Diaz de Haro 3,
E-48013 Bilbao, Spain}
\affiliation{Donostia International Physics Center (DIPC), Paseo Manuel de
  Lardizabal 4, E-20018 San Sebasti\'{a}n, Spain}

\author{David Jacob}\email{david.jacob@ehu.es}
\affiliation{Nano-Bio Spectroscopy Group and European Theoretical Spectroscopy
Facility (ETSF), Dpto. de F\'{i}sica de Materiales,
Universidad del Pa\'{i}s Vasco UPV/EHU, Av. Tolosa 72,
E-20018 San Sebasti\'{a}n, Spain}
\affiliation{IKERBASQUE, Basque Foundation for Science, Maria Diaz de Haro 3,
E-48013 Bilbao, Spain}

\begin{abstract}
  By reverse-engineering from exact solutions we obtain
  Hartree-exchange-correlation (Hxc) potentials for a double 
  quantum dot subject to generic density-density interactions
  and Hund's rule coupling. We find ubiquitous step structures 
  of the Hxc potentials that can be understood and derived from 
  an analysis of stability diagrams. We further show that a 
  generic Hxc potential can be decomposed into four basic potentials which allows
  for a straight-forward parametrization and paves the road 
  for the construction of Hxc potentials
  for interacting multi-orbital systems. 
  Finally we employ our parametrization of the Hxc potential in density
  functional theory calculations of multi-orbital quantum dots and find
  excellent agreement with exact many-body calculations.
\end{abstract}

\date{\today}

\maketitle

\section{Introduction}
\label{sec:intro}

Density functional theory (DFT) is one of the most successful and popular
approaches for computing the electronic structure of molecules and
solids.\cite{HohenbergKohn:64,Kohn:PR:1965,DreizlerGross:90} 
Its success is largely owed to its relative simplicity as well as its low
computational cost as compared to other quantum many-body approaches.
While DFT is an in principle exact theory for computing the ground state
energy and density of a many-electron system, 
in practice approximations have to be made to the
exchange-correlation (xc) part of the total energy functional. The most
popular approximations are the local density (LDA)\cite{Kohn:PR:1965} and
generalized gradient approximations (GGA)\cite{Perdew:85,Becke:88,PerdewBurkeErnzerhof:96,PBEsol:08}
in condensed matter physics, and the so-called hybrid functionals in chemistry.\cite{Becke:93-2,HSE:03}
While these approximations usually work quite well for systems with weak to
moderate electronic interactions, they completely fail for so-called
strongly correlated systems where the interactions between electrons dominate
over the kinetic energy. Since DFT is a formally exact theory which is valid also in the
strongly correlated regime, this failure has to be assigned to shortcomings of
the approximations used. Apparently, the crucial ingredient missing in
standard functionals is the so-called derivative discontinuity,
\cite{PerdewParrLevyBalduz:82} i.e., the discontinuous jump of the
exact xc potential of an open system as the particle number crosses an 
integer.\cite{SagvoldenPerdew:08,GoriGiorgiSavin:09}
In strongly correlated systems the derivative discontinuity contributes a substantial part,
e.g., to the fundamental gap\cite{ShamSchluter:85} or plays a crucial role in the binding and dissociation of molecules.\cite{PerdewParrLevyBalduz:82}

The essential physics of strongly correlated systems can already be captured
by simple but highly nontrivial lattice models such as
the Anderson\cite{Anderson:61} or the Hubbard model\cite{Hubbard:PRSLA:1963} which can be solved by 
advanced many-body methods as for example 
Dynamical Mean-Field Theory (DMFT).\cite{Georges:RMP:1996} 
One way around the problems of the standard approximations of DFT is then 
to combine DFT with advanced many-body calculations of lattice models.
In these approaches the model Hamiltonian describes the strongly interacting part of 
the system, while the weakly to moderately interacting part is still 
described at the level of DFT. 
An important example is the combination of DFT with DMFT (DFT+DMFT).\cite{Kotliar:RMP:2006}
Originally established for the description of bulk materials,
more recently the DFT+DMFT approach has been applied to the description of
nanoscale systems\cite{Jacob:PRB:2010} and molecules\cite{Weber:PNAS:2014}. 
However, this approach is hampered by the so-called double-counting
problem\cite{Karolak:JESRP:2010}, limiting its predictivity. 
More recently, however, new efforts in combining DFT with lattice models 
avoiding the double-counting problem\cite{Requist:PRB:2019,Coe:PRB:2019,Mazouin:PRB:2019}
or to solve the double-counting problem in DFT+DMFT\cite{Haule:PRL:2015} have been undertaken.

Lattice models can also be solved by a lattice version of DFT, an idea which has been pioneered by
Sch\"onhammer.\cite{GunnarssonSchoenhammer:86,SchoenhammerGunnarssonNoack:95}
Later this approach has been further extended to study lattice problems not
only in equilibrium\cite{LimaOliveiraCapelle:02,LimaSilvaOliveiraCapelle:03,
LopezPastor:03,XianlongPoliniTanatarTosi:06,CapelleCampo:13,BroscoYingLorenzana:13,
YingBroscoLorenzana:14,CarrascalFerrerSmithBurke:15,MuellerToewsPastor:19}
but also in out-of-equilibrium situations such as external time-dependent
driving fields\cite{vsa.2006,Verdozzi:08,kskvg.2010,FuksMaitra:14,FuksMaitra:14-2,
DittmannSplettstoesserHelbig:18,DittmannHelbigKennes:19} 
or (steady-state) transport \cite{StefanucciKurth:11,blbs.2012,tse.2012}.
A common theme in many of these studies is again the crucial role of the
derivative discontinuity (or, alternatively, step features in the xc
potential) in the correct description of strongly correlated systems. For
instance, for the Hubbard model the derivative discontinuity is the only
contribution to the band gap, i.e. the mechanism responsible
for opening the Mott-Hubbard gap within DFT.\cite{LimaOliveiraCapelle:02} 

Here, in the same framework of lattice DFT, we aim for a better
understanding of the structure of the (equilibrium) xc potentials of small
multi-orbital models.
This is motivated by earlier studies on the single-impurity
Anderson model \cite{StefanucciKurth:11,blbs.2012,tse.2012}, i.e. a single
interacting impurity in contact with non-interacting leads. There it was
shown that the Kondo effect, one of the hallmarks of strong correlation
in the context of electron transport, can be described at the DFT level
if the corresponding xc potential exhibits a step at integer particle
number on the impurity. This step is already present in the xc potential
of an isolated impurity in contact with a particle and heat bath described
via the grand canonical ensemble\cite{StefanucciKurth:11}. The
exact xc potential of the impurity connected to leads has the same
qualitative features as the uncontacted one with the quantitative difference
that the broadening of the step is determined (at low temperature) by the
coupling to the leads while in the uncontacted case it is determined by the
temperature of the bath. Therefore, we expect that knowledge of the xc 
potential of multi-orbital models in equilibrium, apart from being 
interesting in itself, have a direct relevance in the context of transport. 

To this end we extensively study a double quantum dot
(DQD) subject to generic interactions as the simplest possible model system
for a strongly correlated multi-orbital system. Using reverse-engineering, we
construct the exact xc potentials whose essential features are step structures
which depend on the particular choice of the interaction. We illustrate how
these step structures can be inferred from an analysis of the stability
diagram, i.e., regions of different ground states in the parameter space given
by the single-particle level structure. Analysis of the reverse-engineered
xc potentials reveals that they can be constructed from four basic building
blocks which can be rationalized by a corresponding decomposition of the
electron-electron interaction. It is exactly this decomposition of the
interaction and the corresponding xc potentials %into basic building blocks
which then allows us to build xc functionals for multi-orbital quantum dots
with more than two orbitals.

\section{Model}
\label{sec:model}

We consider a multi-orbital quantum dot (QD) with $\mathcal{M}$ orbitals 
subject to direct Coulomb repulsion and Hund's rule coupling. 
The corresponding Hamiltonian reads:
\begin{eqnarray}
  \label{eq:H}
  \mathcal{H} &=& \sum_{\alpha}\, v_\alpha \hat{n}_\alpha + \sum_{\alpha} U_\alpha\,\hat{n}_{\alpha\su} \hat{n}_{\alpha\sd}
                  + \sum_{\alpha<\beta} U_{\alpha\beta} \, \hat{n}_\alpha \hat{n}_\beta
                  \nonumber\\
              &-& \sum_{\alpha<\beta,\sigma} J_{\alpha\beta} \, \left[ \hat{n}_{\alpha\sigma} \hat{n}_{\beta\sigma}
                  + \left( c^\dagger_{\alpha\sigma} c_{\alpha\bar\sigma}\, c^\dagger_{\beta\bar\sigma} c_{\beta\sigma} \right)  \right]
\end{eqnarray}
where $c_{\alpha\sigma}$ ($c_{\alpha\sigma}^\dagger$) are the annihilation (creation) operators for orbital $\alpha$ and spin $\sigma$, 
$\hat{n}_{\alpha\sigma}$ is the corresponding number operator and $\hat{n}_\alpha=\hat{n}_{\alpha\su}+\hat{n}_{\alpha\sd}$.
$U_\alpha$ is the direct intra-orbital Coulomb repulsion for orbital $\alpha$, 
$U_\alpha\equiv\bra{\alpha,\alpha}\hat{\mathcal{V}}_c\ket{\alpha,\alpha}$, and $U_{\alpha\beta}$ is the direct inter-orbital 
Coulomb repulsion between electrons in two different orbitals, $U_{\alpha\beta}=\bra{\alpha,\beta}\hat{\mathcal{V}}_c\ket{\alpha,\beta}$ 
for $\alpha\ne\beta$. $J_{\alpha\beta}$ is the Hund's rule coupling, i.e. 
the exchange integral of the Coulomb interaction, $J_{\alpha\beta}=\bra{\alpha,\beta}\hat{\mathcal{V}}_c\ket{\beta,\alpha}$ for $\alpha\ne\beta$.
Note that here we have split already the Hund's rule term into the density-density contribution (first term in the last line) and the spin-flip contrbitution (last term).
$v_\alpha$ are the orbital energies (single-particle energies) of the orbitals $\alpha$ which can be tuned by an external ``gate'' 
potential. From here on we will thus refer to $v_\alpha$ as the gate potential or simply gate for orbital $\alpha$.

Here we work at (typically small) finite temperature $T$ and consider the grand
canonical ensemble (GCE). The corresponding density matrix (statistical
operator) is thus given by
\begin{equation}
  \hat\Gamma = \frac{e^{-\beta \mathcal{H}}}{Z} = \frac{1}{Z}\sum_m e^{-\beta E_m} \ket{m}\bra{m}
\end{equation}
where $\beta=1/T$ and $Z$ is the GCE partition function with the chemical
potential $\mu$ set to zero for convenience. The $\ket{m}$ are the many-body
eigenstates of the QD and $E_m$ the corresponding eigenenergies, i.e.
$\mathcal{H} \ket{m} = E_m \ket{m}$. Note that in the absence of Hund's rule coupling ($J_{\alpha\beta}=0$) 
the many-body eigenstates $\ket{m}$ are simply Slater determinants built from the single particle orbitals 
$\ket{\phi_{\alpha\sigma}}=c_{\alpha\sigma}^\dagger\ket{0}$ where $\ket{0}$ is the vacuum state.

The Hamiltonian (\ref{eq:H}) is very common in the fields of strongly correlated electrons
and mesoscopic physics, as it provides a natural description of 3d- or 4f-shells of transition 
metal or lanthanide impurities in metallic hosts and of multi-orbital quantum dots. In these
systems density-density interactions and Hund's rule coupling are by far the most important
interactions. In particular the role of the latter has become a focus of intense research 
in the field of strongly correlated electrons in recent years.\cite{Georges:12} 
Alternatively, the Hamiltonian (\ref{eq:H}) may be viewed as a lattice Hamiltonian in the 
limit of vanishing hopping between sites which is similar to the point of view taken in
strictly correlated DFT.\cite{Seidl:99,Mirtschink:13}  

\section{Reverse-engineering of Hxc potentials at finite temperature}
\label{sec:reverse}

In our truncated Hilbert space the density is uniquely defined by all occupancies 
${\bm n}\equiv(n_1,\ldots,n_{\mathcal{M}})$ of the QD orbitals.
The Mermin theorem\cite{Mermin:65} (the finite-temperature version of the
Hohenberg-Kohn theorem\cite{HohenbergKohn:64}) then establishes 
a one-to-one correspondence between the density (occupancies) ${\bm n}$ and 
the external potential (gate) ${\bm v}\equiv(v_1,\ldots,v_{\mathcal{M}})$: 
${\bm n}\stackrel{1-1}{\longleftrightarrow}{\bm v}$
or, in other words, the external potential is a functional of the occupancies, i.e., 
$v_\alpha=v_\alpha[{\bm n}]$. 

In order to proceed we introduce the Kohn-Sham (KS) system 
i.e. an effective non-interacting system that exactly reproduces the
density ${\bm n}$ of the many-body Hamiltonian $\mathcal{H}$.\cite{Kohn:PR:1965}
Here the KS Hamiltonian is already diagonal in the original single-particle basis, i.e.,
\begin{equation}
  \mathcal{H}^s = \sum_\alpha v_\alpha^s \, \hat{n}_\alpha.
\end{equation}
and the KS orbitals are identical to the original basis orbitals $\ket{\phi^s_{\alpha\sigma}}\equiv\ket{\phi_{\alpha\sigma}}$
with their eigenenergies given by the KS (gate) potentials $v^s_\alpha$.
The Hartree-exchange-correlation (Hxc) potentials $v^\Hxc_\alpha$ are
defined as the difference between the KS gate and the actual gate potential:
\begin{eqnarray}
  \label{eq:vHxc}
  v^\Hxc_\alpha[ {\bm n}] &=& v^s_\alpha[{\bm n}]
  - v_\alpha [{\bm n}].
\end{eqnarray}
The Hxc potential depends on the electron density which is completley determined by the
occupancies of the QD orbitals $\bm n$.

In order to determine the Hxc potential ${\bm v}^\Hxc$ as a functional of the density $\bm n$,
the many-body problem given by $\mathcal{H}$ is solved for a given set of gates
$\bm v$. The resulting set of eigenstates and corresponding energies determines
the density in the GCE according to:
\begin{equation}
  \label{eq:occ}
  n_\alpha = \Tr[ \hat\Gamma \,\hat{n}_\alpha ] = \frac{1}{Z}\sum_m \bra{m} \hat{n}_\alpha \ket{m} e^{-\beta E_m}.
\end{equation}
The density in turn uniquely determines the KS potential and thus the Hxc potential.
In our case of an isolated QD at finite temperature the occupancy $n_\alpha$ 
is simply determined by the gate $v^s_\alpha$ of a non-interacting QD,
and is thus simply given by the Fermi-Dirac distribution, i.e. $n_\alpha = 2\,f(v^s_\alpha)$.
Hence the KS gate for orbital $\alpha$ is given by $v^s_\alpha = \frac{1}{\beta}\ln\left(\frac{2}{n_\alpha}-1\right)$
and the corresponding Hxc potential can be obtained using (\ref{eq:vHxc}) as:
\begin{equation}
  \label{eq:vhxc}
  v^\Hxc_\alpha = \frac{1}{\beta}\ln\left(\frac{2}{n_\alpha}-1\right) - v_\alpha.
\end{equation}
Hence we have found the mapping ${\bm n}\longrightarrow{\bm v}^\Hxc$. 
By exploring the parameter space ${\bm v}=(v_1,\ldots,v_{\mathcal M})$ 
we can establish this mapping for the entire space of densities ${\bm n}=(n_1,\ldots,n_{\mathcal M})$ (for $n_\alpha\in[0,2]$).

\subsection{Hxc potentials and link to stability diagrams for the double quantum dot}
\label{sub:Hxc_and_stability}

We now focus on the two-orbital case, i.e. a double quantum dot (DQD) with generic
density-density interactions ($U_1$, $U_2$, $U_{12}$). For the moment we 
neglect Hund's rule coupling ($J_H=0$), but we will discuss the effect of
finite $J_H$ later in Sec.~\ref{sub:hund}. The Hxc potentials can be constructed
by reverse engineering as explained above. Here we are interested in the
qualitative structure of the Hxc potentials, in particular in the positions
(and heights) of step structures which appear in the low-temperature limit.
In fact, these steps are not only the crucial but also the {\em only} features
of the Hxc potential in the limit of low temperatures. In this section we will
show how these step structures can be deduced completely from the stability
diagrams.

A stability diagram highlights the occupations (densities) of the ground
states in the different regions of the plane of external gates 
$v_1$ and $v_2$. The position and shape of these regions in the $v_1$-$v_2$
plane depend on the values of the interaction parameters but within each
region the pair of densities $(n_1,n_2)$ remains constant at (close to) zero
temperature and the possible values of the local densities $n_i$ are
restricted to 0, 1, and 2.
For general temperatures the domain of physically realizable
densities is restricted to the square $0\leq n_i\leq2$. In this domain
of densities, each of the nine pairs of densities $(n_1,n_2)$ with
$n_i \in \{0,1,2\}$ corresponds to a single point which we call a vertex.
Therefore a region of constant density in the stability diagram (i.e., in
the $v_1$-$v_2$ plane) directly corresponds to a vertex (a single point) in
the domain of realizable densities. A similar duality between regions in the plane 
of ``potentials'' and vertices in the plane of ``densities'' has also been observed
in the framework of steady-state DFT (i-DFT).\cite{StefanucciKurth:15}

It turns out that the structure of the Hxc potentials (in the limit of low
temperatures) for a given set of interaction parameters can be extracted just
by looking at the stability diagram: for a given pair of ground state
densities (or vertex in the density domain) one just needs to
find all adjacent regions corresponding to a different vertex. Then the
Hxc potentials, which are functions of the density, will
{\em only} contain steps which connect a given vertex
(in the density domain) with those vertices
corresponding to directly adjacent regions. The heights of these steps can
also be extracted from the stability diagram. Below we will illustrate how this
works presenting some representative examples and we will also explain the
physical reasons behind our observations.

\begin{figure}[t]  
  \includegraphics[width=\linewidth]{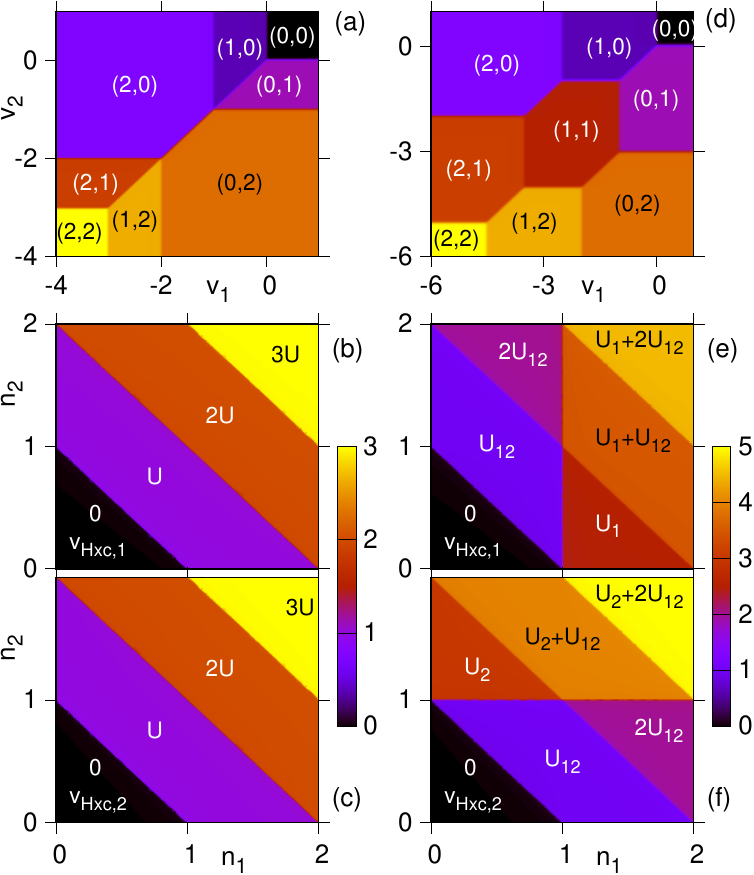} 
  \caption{Panels (a)-(c) (constant interaction model, CIM): stability diagram
    (a) and Hxc potentials for orbitals 1 and 2 (panels (b) and (c),
    respectively) of the double quantum dot for $U_1=U_2=U_{12}$.
    Panels (d)-(f) (Regime I): stability diagram (d) and Hxc potentials for
    orbitals 1 and 2 (panels (e) and (f), respectively) of the double quantum dot
    for $U_1=2.5U_{12}$, $U_2=3U_{12}$. All energies in units of smallest interaction ($U_{12}$).
  }
\label{vhxc_cim_reg1}
\end{figure}

As a first example we choose a simple one where all interaction parameters
are equal, $U_1=U_2=U_{12}$. In this case the total interaction can be written as
$\frac{1}{2} U \hat{N}(\hat{N}-1)$ where $\hat{N}=\hat{n}_1+\hat{n}_2$ is the
operator for the total number of electrons on the dot. This model is known as
the constant interaction model (CIM). It can be shown
\cite{StefanucciKurth:13} that at zero temperature the Hxc potential
$v^{\rm Hxc}_{\alpha}$ of the CIM is independent of $\alpha$ and is a piecewise
constant function of the total electron number $N$ with discontinuous steps
of height $U$ whenever $N$ crosses an integer. We mention that the CIM Hxc
potential is strictly discontinuous only at zero temperature (this is a
manifestation of the famous derivative discontinuity of DFT
\cite{PerdewParrLevyBalduz:82}). At finite but small temperature, the step
structure persists but the Hxc potentials are now continuous functions of the
densities\cite{KurthStefanucci:17}.

We now show how the known CIM Hxc potentials (at low temperature) can be
inferred directly from the stability
diagram. This diagram is shown in panel (a) of Fig.~\ref{vhxc_cim_reg1} for
the CIM with $U_1=U_2=U_{12}=1$. Here the regions corresponding to the
different possible ground state densities (given in parenthesis) are marked by
different colors. The reverse-engineered Hxc potentials for orbitals 1 and 2 are
shown in panels (b) and (c) of Fig.~\ref{vhxc_cim_reg1}, respectively. 
In the stability diagram, the domain corresponding to the occupation
$(0,0)$ is directly adjacent only to the domains with occupations
$(1,0)$ and $(0,1)$. If we connect the $(0,0)$ vertex with one of those
vertices in the $n_1$-$n_2$ plane we see that the resulting lines run along the
border of the allowed density domain. The complete set of lines along the
borders of the density domain follow from the sequence of vertices
$(0,0)\to(1,0)\to(2,0)$, $(0,0)\to(0,1)\to(0,2)$, $(2,0)\to(2,1)\to(2,2)$,
and $(0,2)\to(1,2)\to(2,2)$. The only other possibilities of connecting
vertices corresponding to adjacent regions in the $v_1-v_2$ plane are
(i) $(1,0) \to (0,1)$, (ii) $(2,0) \to (0,2)$, and (iii) $(2,1) \to (1,2)$.
These lines are exactly the position of the steps at integer $N=n_1+n_2$
in the Hxc potentials, see panels (b) and (c) of Fig.~\ref{vhxc_cim_reg1}.
Moreover, the height of these steps can also be deduced from the stability
diagram: consider a $v_2>0$ such that the second dot is always empty,
independent of $v_1$. Then we have essentially a single-site model (SSM) 
in contact with a particle and heat bath because
the second dot doesn't contribute. However, we know that the
the Hxc potential of a SSM in the low temperature limit is a step
function with step of height $U$ at half filling\cite{StefanucciKurth:11}.
In a self-consistent DFT calculation this step in the Hxc potential leads
to a pinning of the KS level to the Fermi energy over a range of gates
of width $U$. Therefore, the width (in $v_1$) of the $(1,0)$ region is just
$U_1=U$. Similarly, the width (in $v_2$) of the $(0,1)$ region is $U_2=U$.
The line in the stability diagram where the $(1,0)$ and $(0,1)$ regions touch
is the line $v_1=v_2$ for which the states with the corresponding
occupations are degenerate.
The KS system is a system of effectively non-interacting electrons which
reproduces the interacting density. However, for a non-interacting double dot
with potentials $v^{s}_{1}$ and $v^{s}_{2}$ the {\em only} possibility for
the half-filled dots to be degenerate is for the point
$v^{s}_{1}=v^{s}_{2}=\ln{3}/\beta$ which in the limit of zero temperature
approaches $v^{s}_{1}=v^{s}_{2}=0$.
Therefore, in order to reproduce the degeneracy as observed in the stability 
diagram, the KS potential for {\em both} orbitals has to be pinned over
an interval of range $U$. This can only be achieved if the Hxc potentials
for both orbitals have steps at $N=1$ of height $U$ as observed in the
reverse-engineered Hxc potentials. On the line $v_1=v_2$ also the states
with occupations $(2,0)$ and $(0,2)$ as well as the ones with
occupations $(2,1)$ and $(1,2)$ are degenerate. It is easy to show that
along this line the states with occupations $(2,0)$ and $(0,2)$ are
lowest in energy for the region $-U> v_1 > -2 U$. 
For non-interacting systems, again there is only one point ($v^{s}_{1}=v^{s}_{2}=0$)
for which the states with $(2,0)$ and $(0,2)$ are degenerate. Therefore,
the Hxc potentials have to be such that for the range of gates
$-U> v_1 > -2 U$, the KS potentials are pinned to zero. This can only be
achieved if both Hxc potentials exhibit a step of height $U$ at
$N=2$, as observed. Finally, there is yet another step of height $U$ in both
Hxc potentials for $N=3$ which follows in a similar way from the analysis of
the contact line between the $(2,1)$ and $(1,2)$ regions. In this way we have
therefore been able to reconstruct the Hxc potentials of the CIM only
by analyzing the stability diagram.

In the second example we make all the interaction parameters different from
each other, i.e., the levels are now not equivalent any more. Moreover, we
choose the interdot interaction $U_{12}$ to be smaller than both $U_1$ and
$U_2$. 
This parameter regime ($U_{12}<U_1,U_2$) we denote as Regime I, see discussion in Section
\ref{sub:decomposition}. In the stability diagram for this regime (panel (d)
of Fig.~\ref{vhxc_cim_reg1}) we now have a new region with densities $(1,1)$
showing up. In order to deduce the low-temperature Hxc potentials
(reverse-engineered results shown in panels (e) and (f) of
Fig.~\ref{vhxc_cim_reg1}), we begin by looking at the regions with
occupations $(1,0)$ and $(0,1)$. The corresponding states are degenerate
along the line $v_1=v_2$ and for $-U_{12}<v_1<0$ they are the ground states of
the double dot. For the KS system to reproduce this density for external
potentials $v_1=v_2$ in the same interval, we need the KS potentials on both
orbitals to be pinned to the Fermi energy. Therefore both Hxc potentials need
to exhibit a step of height $U_{12}$ along the line connecting the vertices 
$(1,0)$ and $(0,1)$. If one of the levels is completely empty, the other
level essentially behaves like a SSM (see discussion of the previous example).
Therefore, for the Hxc potential of orbital 1 we have $v^{\rm Hxc}_1(n_1,0) = U_1$ 
for $1<n_1<2$ while $v^{\rm Hxc}_2(0,n_2) = U_2$ for $1<n_2<2$. The
regions $(1,0)$ and $(1,1)$ are adjacent along the line $v_1=-U_{12}$ for
$-U_1<v_1<-U_{12}$ and thus the KS potential of the first orbital needs
to be pinned to the Fermi energy for this range of $v_1$ leading to a step
of height $U_1-U_{12}$ along the line connecting the $(1,0)$ and $(1,1)$
vertices for $v^{\rm Hxc}_1$. Similarly, $v^{\rm Hxc}_2$ needs to exhibit a step
of height $U_2-U_{12}$ along the line connecting the $(0,1)$ and $(1,1)$
vertices. Next, the regions $(2,0)$ and
$(1,1)$ are adjacent for $-U_{12}-U_{\alpha} < v_{\alpha} < -U_{\alpha}$
($\alpha=1,2$) and therefore both Hxc potentials have a step of height
$U_{12}$ along the lines connecting the $(2,0)$ and $(1,1)$ vertices.
Similarly, there also has to be a step of height $U_{12}$ in both Hxc
potentials along the line connecting the $(0,2)$ and $(1,1)$ vertices.
The regions $(1,1)$ and $(1,2)$ are adjacent for $- 2 U_{12}< v_1<-U_1-U_{12}$
leading to a step of height $U_1-U_{12}$ in $v^{\rm Hxc}_{ 1}$
along the line $(1,1) \to (1,2)$.
Similarly, there is a step of height $U_1-U_{12}$ in $v^{\rm Hxc}_{2}$ along the
$(1,1) \to (2,1)$ line. Finally, the regions
$(2,1)$ and $(1,2)$ are adjacent along a line of length $U_{12}$ leading to a
step of this height in both Hxc potentials along the $(2,1) \to (1,2)$ line.
In this way we now have completely determined
the (low temperature) Hxc potentials of both orbitals just by analyzing the
stability diagram. Their overall structure is such that they exhibit steps
for integer total occupation
$N=n_1+n_2$ for both Hxc potentials plus an additional step at
$n_{\alpha}=1$ for $v^{\rm Hxc}_{\alpha}$. Note also that for the
special case $U_{12}=0$ only the steps at $n_{\alpha}=1$ for $v^{\rm Hxc}_{\alpha}$
survive while those at integer $N$ disappear. This is not surprising
since in this case our model just describes two completely independent
single impurities and, naturally, the corresponding Hxc potential for orbital
$\alpha$ is completeley independent of the other orbital and given by the
Hxc potential of a SSM with interaction strength
$U_{\alpha}$. This has also been discussed as ``intra-system steps'' in
Ref.~\onlinecite{DimitrovAppelFuksRubio:16}. 

\begin{figure}[t]  
  \includegraphics[width=\linewidth]{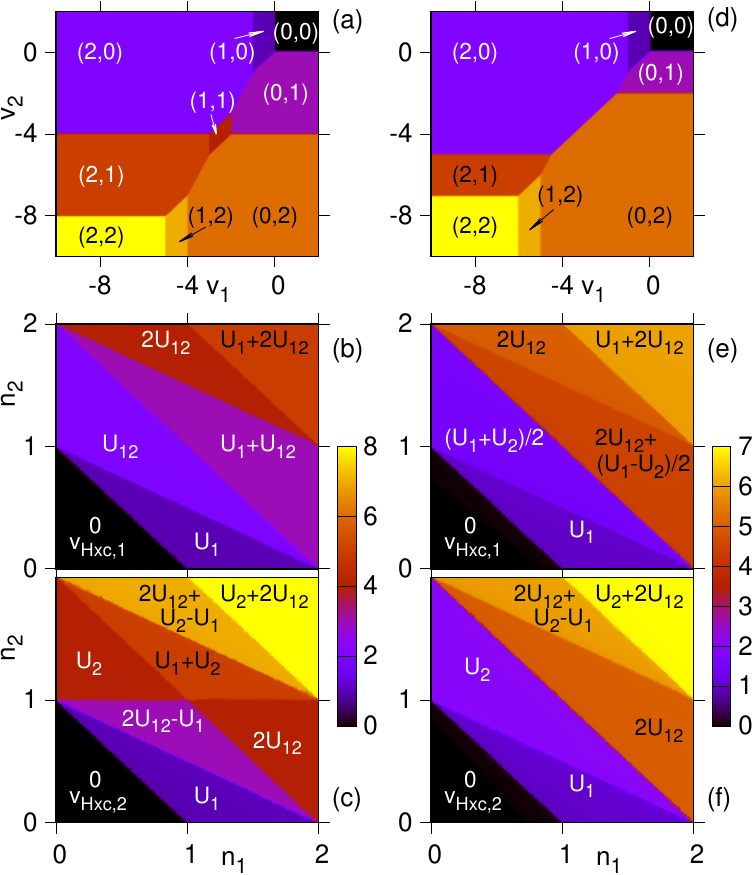} 
  \caption{Panels (a)-(c) (Regime II): stability diagram (a) and Hxc
    potentials for orbitals 1 and 2 (panels (b) and (c), respectively) of the
    double quantum dot for $U_2=4U_1$ and $U_{12}=2U_1$.
    Panels (d)-(f) (Regime III): stability diagram (d) and Hxc potentials for
    orbitals 1 and 2 (panels (e) and (f), respectively) of the double quantum
    dot for $U_2=2U_1$, and $U_{12}=2.5U_1$.
    All energies in units of the smallest interaction ($U_1$).
  }
\label{vhxc_reg2_reg3}
\end{figure}

We have identified two further parameter regimes for the interaction
parameters (see Section \ref{sub:decomposition}) where qualitative changes
both in the stability diagram as well as in the Hxc potentials occur. 
In both regimes the inter-orbital interaction $U_{12}$ is
smaller than at least one of the intra-orbital ones. Without loss of
generality we may assume $U_1\leq U_2$. In Regime II we have $U_1<U_{12}<(U_1+U_2)/2$
while Regime III is defined by $U_1\leq (U_1+U_2)/2\leq U_{12}$.
In panels (a)-(c) of
Fig.~\ref{vhxc_reg2_reg3} we show the stability diagram and Hxc potentials
for interaction parameters chosen in Regime II. Compared to Regime I
(panels (d)-(f) of Fig.~\ref{vhxc_cim_reg1}), in the stability diagram we
now find that there exists a range of potentials for which
regions $(2,0)$ and $(0,1)$ are directly adjacent and, similarly, for the 
regions $(2,1)$ and $(0,2)$. As expected, these transitions lead to
the new steps in the Hxc potentials. On the other hand, for
the Hxc potential of orbital 1 the step at $n_1=1$ (present in Regime I) now
disappears while in $v_{\rm Hxc,2}$ the step at $n_2=1$ survives (this step is
related to the vertical lines delimiting the $(1,1)$ region in the stability
diagram). We have annotated the plateau values in both Hxc potentials which
can be found by analyzing the stability diagram using similar arguments to
the ones used above for Regime I. 

Finally, in panels (d)-(f) of Fig.~\ref{vhxc_reg2_reg3} we show the stability
diagram and Hxc potentials for interaction parameters chosen in Regime III.
Compared to Regime II, the main qualitative difference is the disappearance
of the step at $n_2=1$ in the Hxc potential of orbital 2. Again, all the step 
structures in the Hxc potentials can fully be deduced by analyzing the
stability diagram. 

Before we close this section we would like to mention that for
multilevel dots beyond the double dot studied here, in principle
the analysis of the (multidimensional) stability diagram(s) also allows
for a complete deduction of the low-temperature Hxc potentials of the
different orbitals. However, it is clear that this procedure rapidly becomes
quite cumbersome as the number of levels increases.

\section{Modelling of the Hxc potentials}
\label{sec:hxc_model}

\subsection{Decomposition of the interaction into basic building blocks}
\label{sub:decomposition}

In the following we show that the Hxc potentials for generic density-density interactions can be built from a 
few basic potentials. We start with the most common (or natural) situation where the inter-orbital interaction $U_{12}$ 
is smaller than both of the intra-orbital ones, $U_{12} \le U_1,U_2$. 
A specific case with $U_{12}<U_1<U_2$ was studied in the previous section [see Fig.~\ref{vhxc_cim_reg1}(d-f)]. 
The corresponding Hxc potential shows steps at integer values of $N=n_1+n_2$, connected to a CIM potential,
as well as steps at $n_1=1$ for orbital 1 or at $n_2=1$ for orbital 2 connected to a SSM potential of the 
corresponding orbital. 

This suggests that in the regime $U_{12} \le U_1,U_2$ the Hxc potential for each orbital may be built from a superposition 
of a CIM potential plus a SSM potential. We can rationalize this idea by a decomposition of the Coulomb interaction term as 
follows. Rewriting the inter-orbital repulsion as
\begin{equation}
  U_{12}\, \hat{n}_1 \, \hat{n}_2 =  \frac{U_{12}}{2}  \hat{N}(\hat{N}-1) - U_{12} \sum_\alpha \hat{n}_{\alpha\su} \hat{n}_{\alpha\sd}
\end{equation}
we can split the interaction into a CIM part and two SSM interactions (one for each orbital):
\begin{equation}
  \label{eq:Decomposition:I}
  \mathcal{V}_{\rm int} = \frac{1}{2} U_{12} \, \hat{N}(\hat{N}-1)  + \sum_{\alpha} \delta{U}_\alpha\,\hat{n}_{\alpha\su} \hat{n}_{\alpha\sd}
\end{equation}
where $\delta{U}_\alpha\equiv U_\alpha-U_{12}$ is the ``excess interaction'' for each orbital. 
This suggests to write the Hxc potential for level $\alpha$ for \textbf{Regime I} ($U_{12} \le U_1,U_2$) as 
the sum of the CIM Hxc potential for interaction $U_{12}$ and the SSM potential for $\delta{U}_\alpha$:
\begin{equation}
  \label{eq:Hxc:I}
  v^\Hxc_\alpha[{\bm n}] =  v^\Hxc_\CIM(U_{12})[N] + v^\Hxc_\SSM(\delta{U}_\alpha)[n_\alpha]
\end{equation}
where $N=n_1+n_2$.

\begin{figure}
  \begin{tabular}{lr}
    \includegraphics[width=0.5\linewidth]{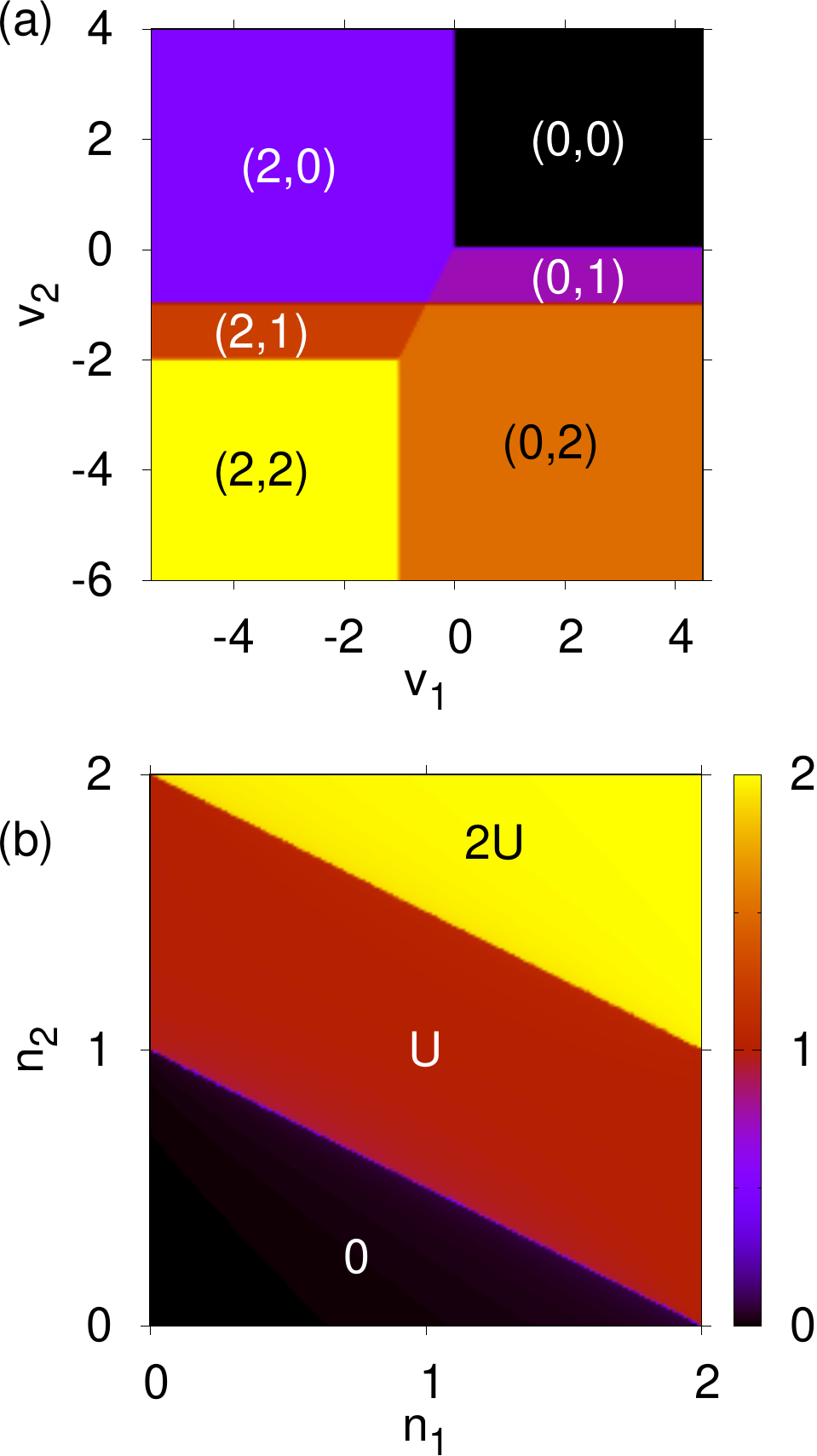} & 
    \includegraphics[width=0.5\linewidth]{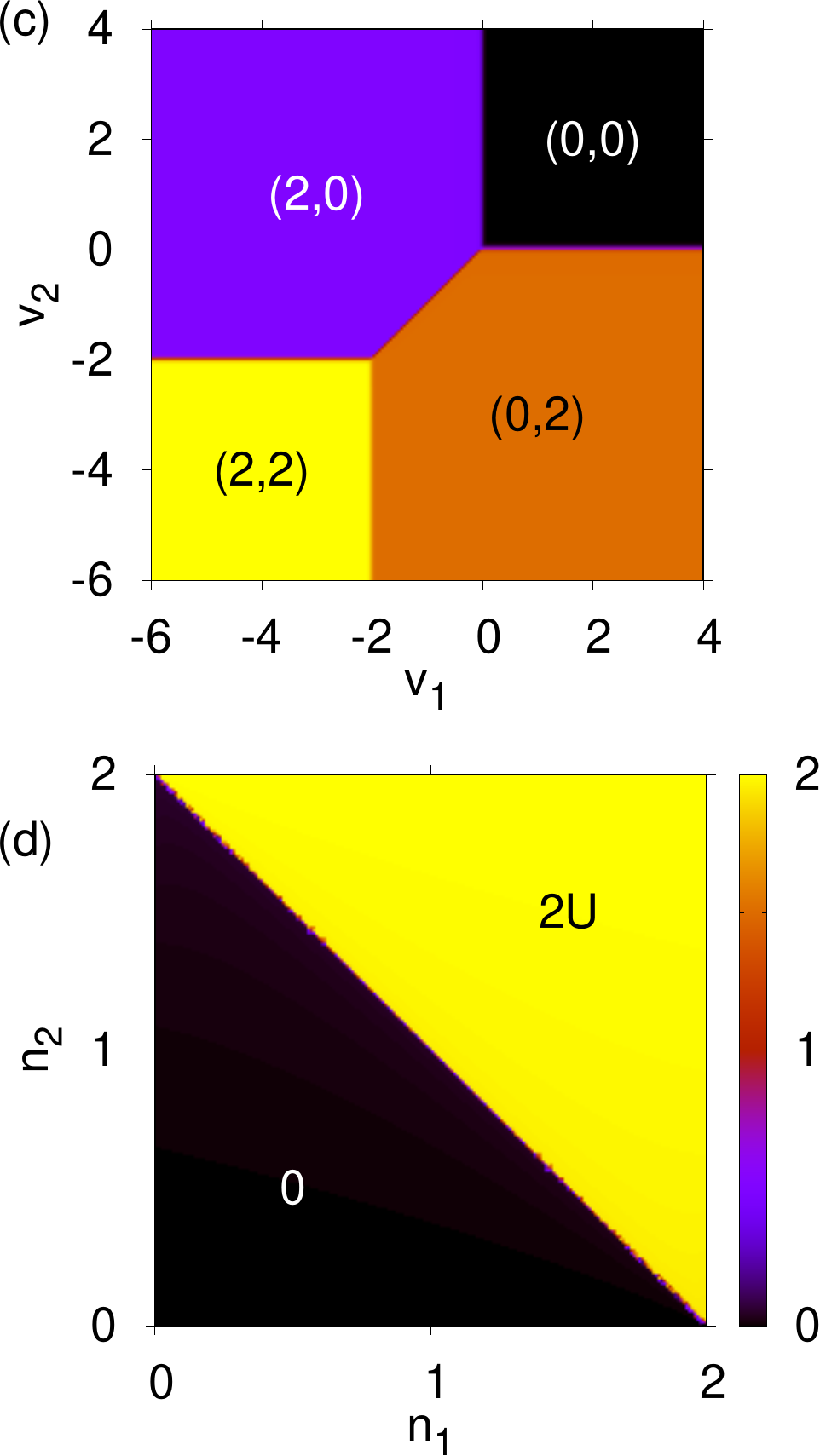} 
  \end{tabular}
  \caption{
    (a,b) Stability diagram (a) and Hxc potential of orbital 2 (b)
    for the Skew interaction $\mathcal{V}_{\ske}=\frac{U}{2}\hat{n}_2(\hat{N}-1)$.
    The structure of the Hxc potential for orbital 1 is the same as for orbital 2
    but the step heights are half those of orbital 2 ($0$, $U/2$, $U$).
    (c,d) Stability diagram (c) and Hxc potential of both orbitals (d) 
    for the inter-orbital interaction $\mathcal{V}_{\inter}=U\hat{n}_1\hat{n}_2$.
    All energies in units of $U$ in both cases. 
  }
  \label{fig:Skew_and_Inter}
\end{figure}

Now if $U_{12}$ is larger than at least one of the intra-orbital interactions $U_\alpha$
this decomposition of the Coulomb interaction obviously leads to negative interactions 
$\delta{U_\alpha}$ in the SSM parts. Since the step in the Hxc
potential of the SSM at $n_\alpha=1$ would actually vanish for negative interactions,\cite{PS.2012}
in this regime the step structure can obviously not be rationalized by the above decomposition
of the interaction.
Indeed the structure of the reverse-engineered Hxc potentials (Fig.~\ref{vhxc_reg2_reg3}) appears to be quite 
different from that for the regime $U_{12}\le U_\alpha$. Essentially, two new features are found
in this regime: (i) an increase of the step height at $N=2$ with respect to the CIM potential, and 
(ii) peculiar new steps at integer values of $n_1/2+n_2$.
The steps at integer $n_1/2+n_2$ are generated by a peculiar interaction of the form
\begin{equation}
  \label{eq:V_skew}
  \mathcal{V}_\ske = \frac{U}{2} \hat{n}_2(\hat{N}-1)
\end{equation}
which we will refer to as \textit{Skew interaction} from now on. 
This interaction is realized by setting $U_1=0$ and $U_{12}=U_2/2=U/2$,
The corresponding stability diagram and the Hxc potential for orbital 2 is shown in Fig.~\ref{fig:Skew_and_Inter}(b). 
Note that the Hxc potential of orbital 1 has the same structure but the step heights are lower by a factor of 1/2. 

Common to all cases is that there is always a contribution of the CIM potential, as long 
as all interactions ($U_1,U_2,U_{12}$) remain finite. This contribution is given by the 
smallest interaction.  In the case that $U_{12}$ is larger than at least one of the intra-orbital
interactions $U_\alpha$, we may assume without loss of generality that $U_1$ is the 
smallest interaction. Subtracting the CIM interaction $\sim{U_1}$ from the total interaction thus 
yields:
\begin{eqnarray}
  \lefteqn{ \mathcal{V}_{\rm int} - \frac{U_1}{2} \hat{N}( \hat{N}-1 ) = }\nonumber\\ 
  &&\hspace{2ex} = (U_{12}-U_1)\, n_1 n_2 + (U_2-U_1) \, n_{2\su} n_{2\sd} 
  \nonumber\\ 
  &&\hspace{2ex} = (U_{12}-U_1)\, n_1 n_2 + \frac{U_2-U_1}{2} \, n_2 ( n_2-1 )
\end{eqnarray}
where in the last term we have rewritten the intra-orbital interaction for orbital 2 
in terms of $n_2=n_{2\su}+n_{2\sd}$ instead of $n_{2\su}$ and $n_{2\sd}$. 
Hence the remaining interaction consists of an inter-orbital interaction $\sim(U_{12}-U_1)$ and a
SSM interaction $\sim(U_2-U_1)/2$ for orbital 2. These two terms can be combined to yield the 
Skew interaction and a remaining term. Depending on whether $U_{12}-U_1$ is larger or smaller
than $(U_2-U_1)/2$, the remaining term is either a SSM interaction for orbital 2 
if $U_{12}-U_1<(U_2-U_1)/2$ (or equivalently $U_{12}<U_\ave\equiv(U_1+U_2)/2$) 
or an inter-orbital interaction if $U_{12}-U_1>(U_2-U_1)/2$ (or equivalently $U_{12}>U_\ave$). 

We thus identify two new regimes in addition to Regime I ($U_{12}\le{U_\alpha}$)
discussed above: In \textbf{Regime II} the inter-orbital interaction $U_{12}$ takes values
between the lowest interaction of both intra-orbital interactions and their average $U_\ave$, 
i.e. $U_1<U_{12}<U_\ave$. This is the case shown in Fig.~2(a-c). 
After subtraction of the CIM potential $\sim{U_1}$ we find that the remaining interaction in Regime II can be written as
\begin{eqnarray}
   (U_{12}-U_1) \, \hat{n}_2 \,(\hat{N}-1) + 2(U_\ave-U_{12})\, \hat{n}_{2\su} \hat{n}_{2\sd}.
\end{eqnarray}
Overall this suggests the following decomposition of the Hxc potential in \textbf{Regime II} ($U_1<U_{12}<U_\ave$):
\begin{eqnarray}
  \label{eq:Hxc:II}
   v^\Hxc_\alpha[{\bm n}] &=& v^\Hxc_\CIM(U_1)[N] \nonumber\\
                          &+&  v^\Hxc_{\ske,\alpha}\left(2(U_{12}-U_1)\right)[{\bm n}] \nonumber\\
                          &+&  v^\Hxc_\SSM\left(2(U_\ave-U_{12})\right)[n_2] \, \delta_{\alpha2}
\end{eqnarray}
where $\delta_{\alpha2}$ is the Kronecker-delta which ensures that the SSM term only contributes to the Hxc potential of orbital 2.
Note that as $U_{12}{\rightarrow}U_\ave$ the SSM term vanishes.

On the other hand \textbf{Regime III} occurs when the inter-orbital interaction exceeds the average intra-orbital 
interaction, i.e. $U_{12}>U_\ave>U_1$. This was the case considered in Fig.~2(d-f).
In this regime the remaining interaction after subtraction of the CIM $\sim{U_1}$ 
can be rewritten in terms of the Skew interaction (\ref{eq:V_skew}) and a pure inter-orbital
interaction part:
\begin{equation}
  \frac{U_2-U_1}{2} \, \hat{n}_2 \,(\hat{N}-1) + (U_{12}-U_\ave)\, \hat{n}_1 \hat{n}_2 .
\end{equation}
As can be seen in Fig.~\ref{fig:Skew_and_Inter}(d), this inter-orbital term
\begin{equation}
  \label{eq:V_inter}
  \mathcal{V}_\inter = U \hat{n}_1 \hat{n}_2
\end{equation}
gives rise to a single step at $N=2$ which explains the increase in step height at $N=2$ with respect to the CIM, observed in Fig.~2(e,f).
Overall this suggests the following decomposition of the Hxc potential in \textbf{Regime~III} ($U_{12}>U_\ave>U_1$):
\begin{eqnarray}
  \label{eq:Hxc:III}
   v^\Hxc_\alpha[{\bm n}] &=& v^\Hxc_\CIM(U_1)[N] \nonumber\\
                          &+& v^\Hxc_{\ske,\alpha}(U_2-U_1)[{\bm n}] \nonumber\\
                          &+& v^\Hxc_\inter(U_{12}-U_\ave)[n_1+n_2] .
\end{eqnarray}
We can see that for $U_1=U_2$ the Skew part of the Hxc potential disappears.

\begin{figure}
  \includegraphics[width=0.9\linewidth]{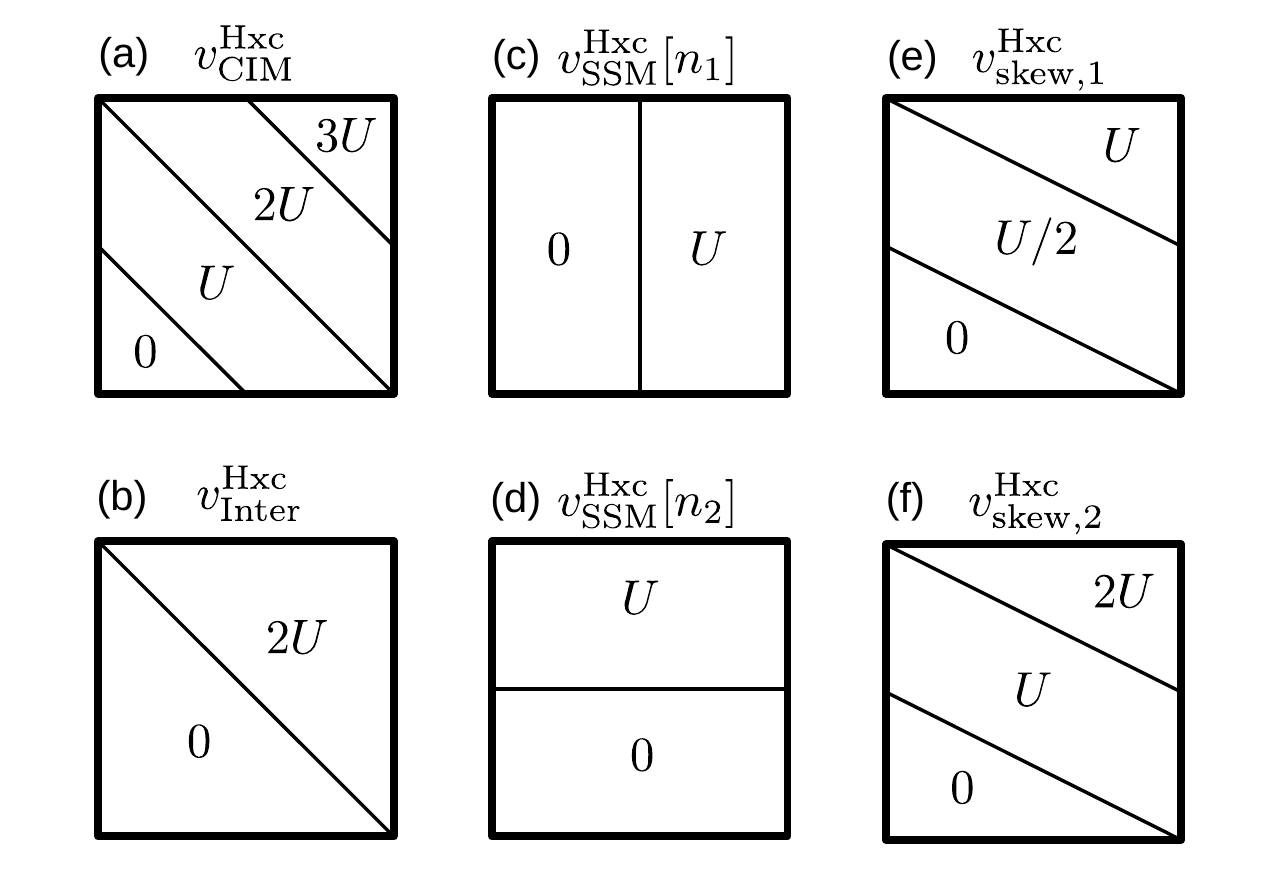}
  \caption{
    Schematic representation of the four basic Hxc potentials for building the generic potentials
    for all three regimes. (a) Hxc potential for CIM interaction $\frac{U}{2}\hat{N}(\hat{N}-1)$. 
    (b) Hxc potential for inter-orbital interaction $Un_1n_2$. (c,d) Hxc potential for intra-orbital 
    (i.e. SSM) interactions $Un_{\alpha\su}n_{\alpha\sd}$. (e,f) Hxc for the Skew interaction 
    $\frac{U}{2}\hat{n}_2(\hat{N}-1)$. 
    \label{fig:basic_potentials}
  }
\end{figure}

Hence we have found a decomposition of the Hxc potential for a two-orbital model with generic (density-density )
interactions in all three regimes in terms of four basic potentials which are shown schematically in Fig.~\ref{fig:basic_potentials}.
We would like to emphasize at this point that \textbf{Regime I} corresponds to a more natural choice of parameters 
than the other two regimes, since the inter-orbital interaction $U_{12}$ is generally smaller than any of the intra-orbital
interactions $U_\alpha$. Nevertheless, the other regimes might be realized by effective models or possibly by screening
of the Coulomb interactions.
In the next section we will present parametrizations of the Hxc potentials in the different regimes, making use of its 
decomposition into the basic building blocks shown in Fig.~\ref{fig:basic_potentials}.

\subsection{Generalization of Hxc potential to more than two orbitals}
\label{sub:mdot_hxc}

For specific choices of parameters we can generalize the Hxc potential for the DQD to an arbitrary number of orbitals 
in a straightforward manner. We concentrate on the physically most relevant \textbf{Regime I} ($U_\alpha,U_\beta>U_{\alpha\beta}$). If we choose the inter-orbital
interaction to be constant, $U_{\alpha\beta}\equiv{U^\prime}$, which thus has to be smaller than \emph{all} of the intra-orbital interactions, $U^\prime<U_\alpha$,
we can rewrite the interaction in a similar manner as in Eq.~(\ref{eq:Decomposition:I}) in terms of a CIM term $\sim{U^\prime}$ for all the electrons $N=\sum_\alpha{n_\alpha}$
and SSM terms $\sim\delta{U}_\alpha\equiv{U_\alpha-U^\prime}$ for the individual orbitals as
\begin{equation}
  \mathcal{V}_{\rm int} = \frac{1}{2}U^\prime \hat{N}(\hat{N}-1) + \sum_{\alpha}\delta{U_{\alpha}} \, \hat{n}_{\alpha\uparrow}\hat{n}_{\alpha\downarrow} 
\end{equation}
where $\delta U_{\alpha}= U_{\alpha} -U'$. This suggests to decompose the XC functionals in complete analogy
to the two-orbital case in Regime I as
\begin{equation}
  \label{eq:Hxc:multi-orbital}
  v^\Hxc_\alpha[{\bm n}] =  v^\Hxc_\CIM(U^\prime)[N] + v^\Hxc_\SSM(\delta{U}_\alpha)[n_\alpha].
\end{equation}
In Sec.~\ref{sub:mdot_results} we will see that this decomposition of the Hxc potential leads
to excellent results for multi-orbital QDs. For a more general choice of interaction 
parameters, the above decomposition is likely to become more complicated. This will be the focus
of future work.

\subsection{The effect of Hund's rule coupling}
\label{sub:hund}

\begin{figure}[t]
  \includegraphics[width=\linewidth]{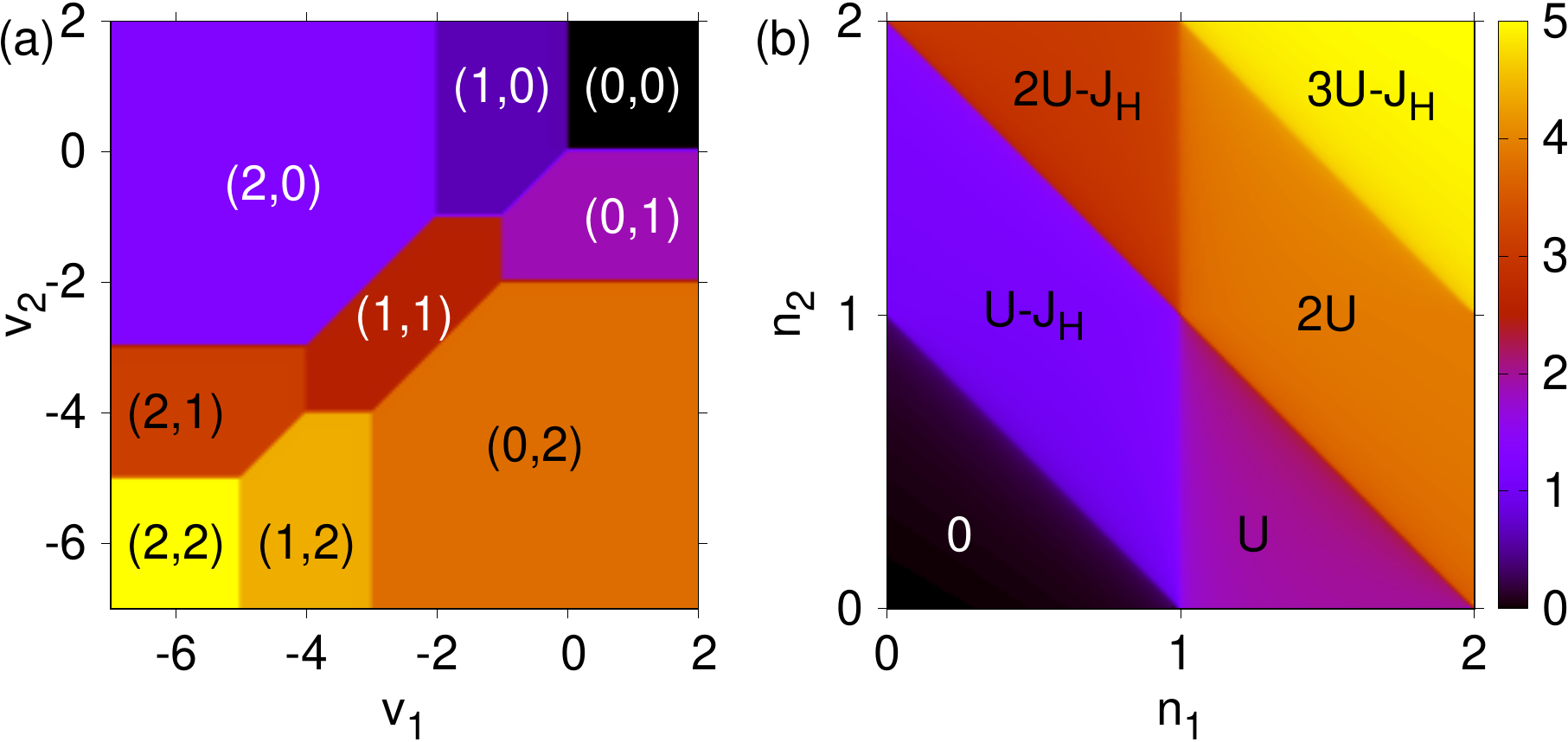}
  \caption{Effect of Hund's rule coupling on (a) Stability diagram and (b) Hxc potential of orbital 1
    for CIM interaction plus Hund's rule coupling, $\mathcal{V}_{\rm int}=\frac{U}{2}\hat{N}(\hat{N}-1)+\mathcal{V}_\Hund$ 
    for $U=2J_H$. Here due to symmetry the Hxc potential for orbital 2 can simply be obtained by
    reflection along the $n_1=n_2$ line. All energies in units of $J_H$. 
  }
\label{fig:Hund}
\end{figure}

So far we have neglected the effect of Hund's rule coupling on the Hxc potentials. In Fig.~\ref{fig:Hund} 
we show the stability diagram and the corresponding reverse-engineered Hxc potential for the case of a
CIM type direct interaction part ($U_1=U_2=U_{12}$) plus the full Hund's coupling contribution ($J_H$). Both
the stability diagram and the reverse-engineered Hxc potential shown in Fig.~\ref{fig:Hund} resemble
the ones for the case with $U_{12}<U_\alpha$ without Hund's coupling (cf. Fig.~\ref{vhxc_cim_reg1}(d-f)).
Only the size of the vertex regions changes in the stability diagram, and correspondingly in the Hxc potentials
only the step heights change.
Moreover, by switching off the spin-flip term in (\ref{eq:H}) we find that it does not have any effect on the densities 
and consequently on the Hxc potentials and thus can be neglected. Hence in the following considerations we only 
need to take into account the density-density part of the Hund's coupling in (next to last term in Eq.~\ref{eq:H}).

In the spirit of the previous section we can rewrite the density-density part of the Hund's rule coupling term 
in terms of a (negative) CIM interaction and (positive) SSM interactions for the remaining orbitals plus a 
remaining \emph{positive} interaction part:
\begin{eqnarray}
  \label{eq:Hund}
  \mathcal{V}_H &=& -J_H \sum_\sigma \hat{n}_{1\sigma} \hat{n}_{2\sigma} = -J_H \hat{n}_1 \hat{n}_2 + J_H \sum_\sigma \hat{n}_{1\sigma} \hat{n}_{2\bar\sigma}
                \nonumber\\
  &=& -\frac{J_H}{2} \hat{N}(\hat{N}-1) + J_H \sum_\alpha \hat{n}_{\alpha\su} \hat{n}_{\alpha\sd} + J_H \sum_\sigma \hat{n}_{1\sigma} \hat{n}_{2\bar\sigma}
      \nonumber\\
\end{eqnarray}
where in the last term $\bar\sigma$ denotes the oposite spin of $\sigma$. The last term gives rise to a
step at $N=2$ of height $J_H$ in the Hxc potential similar to the inter-orbital interaction term but with
step height $J_H$ instead of $2U$ (cf. Fig.~\ref{fig:Skew_and_Inter}(d)).

When adding the density-density contribution of the Hund's rule coupling to the direct interaction part
in Regime I ($U_{12}{\le}U_1,U_2$), we can rewrite the interaction in terms of a CIM interaction, SSM 
terms, and the last term of the Hund density-density interaction (\ref{eq:Hund}) as
\begin{eqnarray}
  \mathcal{V}_{\rm int} &=& \frac{U_{12}-J_H}{2} \hat{N}(\hat{N}-1) + \sum_\alpha(\delta{U}_\alpha+J_H) \hat{n}_{\alpha\su}\hat{n}_{\alpha\sd} \nonumber\\
                    &+& J_H \sum_\sigma \hat{n}_{1\sigma} \hat{n}_{2\bar\sigma}
\end{eqnarray}
where as defined in the previous section $\delta{U_\alpha}=U_\alpha-U_{12}$.
Hence all terms can be modeled by the basic Hxc potentials shown in Fig.~\ref{fig:basic_potentials}:
\begin{eqnarray}
  \label{eq:Hxc:Hund}
   v^\Hxc_\alpha[{\bm n}] &=& v^\Hxc_\CIM(U_{12}-J_H)[N] \nonumber\\
                          &+& v^\Hxc_\SSM(\delta{U}_\alpha+J_H)[{\bm n}] \nonumber\\
                          &+& v^\Hxc_\inter(J_H/2)[n_1+n_2] .
\end{eqnarray}

\subsection{Parametrization of the basic Hxc potentials}
\label{sec:parametrization}

In the zero temperature limit, the steps in the Hxc potential become infinitely sharp and thus
can be described by simple step functions. At finite temperature, however, the steps are broadened
in a non-trivial way. For the SSM at finite temperature an \emph{exact} expression for the Hxc 
potential can be derived:\cite{KurthStefanucci:17}
\begin{equation}
  \label{eq:hxc_ssm}
  v^\Hxc_\SSM[n] = U +  \frac{1}{\beta} \ln\left( \frac{x+\sqrt{x^2+e^{-\beta U}(1-x^2)}}{1+x} \right)
\end{equation}
where $x=n-1$. 

\begin{figure}
    \includegraphics[width=\linewidth]{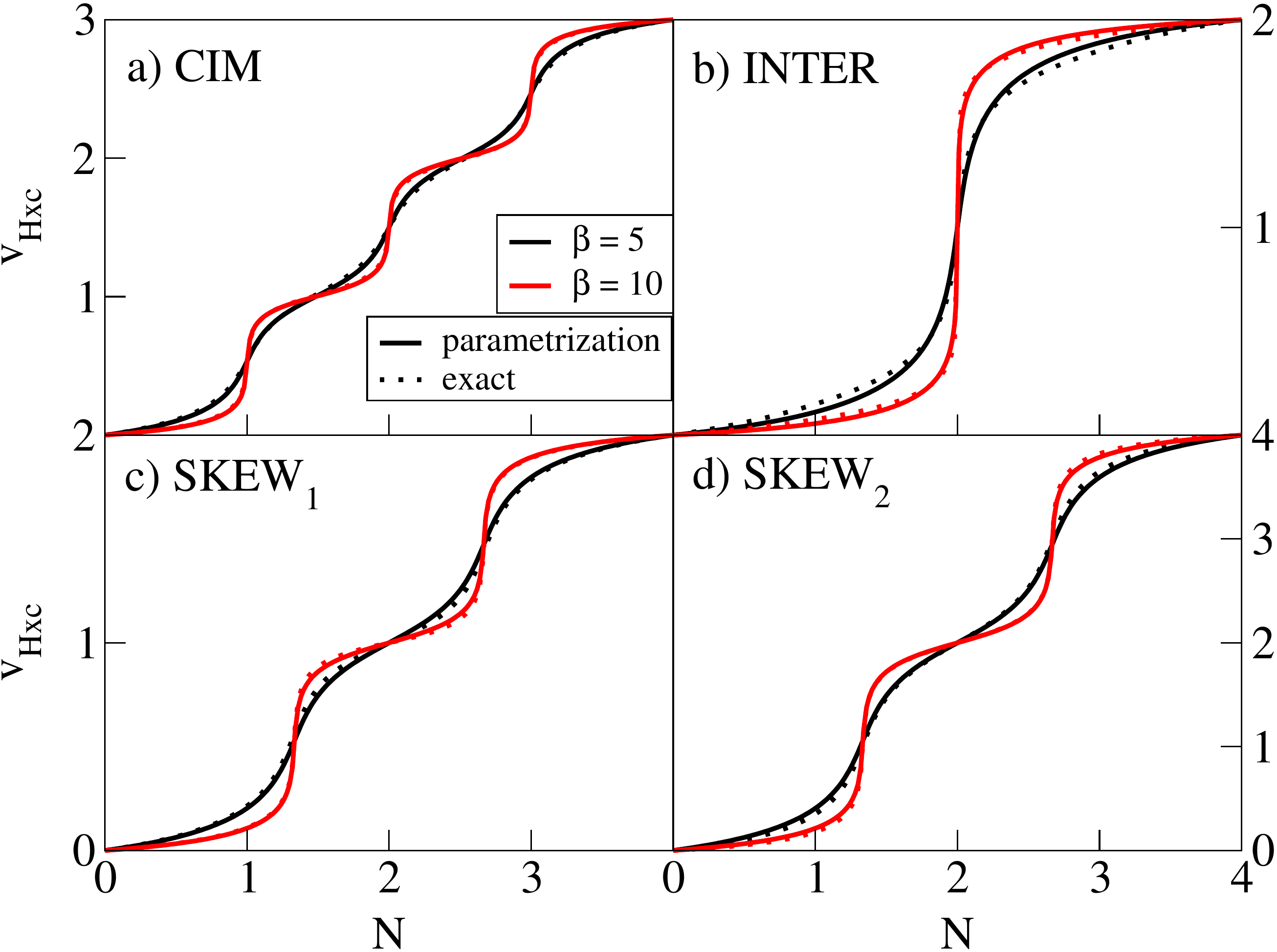} 
  \caption{
    Comparison of parametrized and exact Hxc potentials as a function of $N=n_1+n_2$
    for three basic interactions: (a) CIM interaction ($U_1=U_2=U_{12}>0$); 
    (b) Inter-orbital interaction ($U_{12}>0$ and $U_1=U_2=0$); 
    (c) Skew interaction ($U_2=2U_{12}>0$ and $U_1=0$); 
    All energies in units of the smallest non-zero interaction ($U_{12}$).
    \label{fig:hxc_param}
  }
\end{figure}

In the following we will use the Hxc functional for the SSM as the basis for constructing approximations for
the other three basic Hxc potentials into which the generic Hxc potential can be decomposed, namely the CIM, 
the Inter-orbital, and the Skew potential (see Fig.~\ref{fig:basic_potentials}). 
We start with the CIM potential
and show that an excellent parametrization of the Hxc potential can be achieved by simply summing the (exact) SSM potential (\ref{eq:hxc_ssm})
over the charging states of the dot, and shifting and rescaling it such that the potential does not become negative or larger than $(2\mathcal{M}-1)U$:
\begin{eqnarray}
  \label{eq:hxc_cim}
  \lefteqn{v^\Hxc_\CIM[N] = \frac{(2\mathcal{M}-1)U}{v^{\rm max}_\CIM} } \nonumber\\
  &&\hspace{2ex}\times \sum_{J=1}^{2M-1} \left[ v^\Hxc_\SSM[ N-J+1] - v^\Hxc_\SSM[ -J+1]  \right ]
\end{eqnarray}
where
\begin{equation}
  \label{eq:vmax}
  v^{\rm max}_\CIM = \sum_{J=1}^{2M-1} \left[ v^\Hxc_\SSM[ 2\mathcal{M}-J+1] - v^\Hxc_\SSM[-J+1]  \right ]
\end{equation}
is the maximal value that the sum in (\ref{eq:hxc_cim}) aquires at $N=2\mathcal{M}$. The prefactor $(2\mathcal{M}-1)U/v^{\rm max}_\CIM$ thus
rescales the potential such that the potential yields the exact value $(2M-1)U$ at $N=2\mathcal{M}$. 
As can be seen in Fig.~\ref{fig:hxc_param}(a), the agreement with the exact result is quite remarkable.

For the inter-orbital potential we find a good parametrization describing the step at total $N=2$ again
in terms of the SSM potential, 
as
\begin{equation}
  \label{eq:hxc_inter}
  v^\Hxc_\inter(U,\beta)[N] = v^\Hxc_\SSM(2U,\beta^\ast)[N/2]
\end{equation}
where we have replaced the actual inverse temperature $\beta$ by an effective reduced value, 
$\beta^\ast=0.73\beta$ and the step height is increased by a factor of 2 compared to the SSM.
The agreement with the exact potential is very good as can be seen in Fig.~\ref{fig:hxc_param}(b).

Finally, for the Skew interaction, we parametrize the Hxc potential in a similar way as the 
Hxc potential for the CIM, by summing two SSM potentials, one for each of the steps, and shifting
and rescaling so that the potential does not become negative or larger than the maximum value:
\begin{eqnarray}
  \label{eq:hxc_skew}
  v^\Hxc_{\ske,\alpha}(U)[{\bm n}] &=& \frac{\alpha\,U}{v^{\rm max}_\ske} \sum_{J=0,1} \left\{ v^\Hxc_\SSM(\tfrac{U}{2})\left[\tfrac{n_1}{2}+n_2-J\right] \right. 
  \nonumber\\
  && \hspace{7ex} \left. - v^\Hxc_\SSM(\tfrac{U}{2})[-J]  \right\}
\end{eqnarray}
where 
\begin{equation}
  \label{eq:vmax_skew}
  v^{\rm max}_\ske = \sum_{J=0,1} \left\{ v^\Hxc_\SSM(\tfrac{U}{2})[3-J] - v^\Hxc_\SSM(\tfrac{U}{2})[-J]  \right\} .
\end{equation}
Also here the agreement with the exact potential is very good as can be seen in Figs.~\ref{fig:hxc_param}(c+d).

We have thus found parametrizations of the four basic Hxc potentials.
It should be noted, however, that at higher temperatures the exact CIM and Inter-orbital potentials
(which in the zero temperature limit only depend on total $N$) acquire also a dependence on the difference 
$\delta{N}\equiv{n_1-n_2}$ which has not been taken into account here. This discrepancy of our parametrizations 
might be responsible for some of the moderate deviations of our DFT results from the exact ones discussed in the next section.

\section{DFT calculations}

We are now going to apply the above developed parametrization of the Hxc potential in actual DFT calculations. To this end we solve the 
KS equations which for the multi-orbital QD in the GCE are given by:
\begin{equation}
  \label{eq:ks}
  n_\alpha = 2\,f( v_\alpha+v_\Hxc[{\bm n}] ) \mbox{ for } \alpha=1\ldots\mathcal{M} .
\end{equation}
Since the sharp step features in the Hxc potentials are expected to prevent the convergence
of the usual self-consistency procedure in the density,\cite{xctk.2012} here we make use 
instead of a multidimensional generalization of the bisection approach as proposed before in Ref.~\onlinecite{xctk.2012} for finding the 
root of the multidimensional function 
\begin{equation}
  F_\alpha[{\bm n}]{\equiv}n_\alpha - 2\,f( v_\alpha+v_\Hxc[{\bm n}] ).  
\end{equation}

In the following we study the evolution of the density ${\bm n}$ of multi-orbital QDs 
as a function of the applied gate $v_g$ for different parameter sets. The gate $v_g$ exerts 
a total shift of the QD levels $\epsilon_\alpha$ and hence the total gate for orbital $\alpha$ is given by
\begin{equation}
  v_\alpha = \epsilon_\alpha + v_g .
\end{equation}
Consequently, the differences in the gate potentials between different orbitals
remain constant as the gate $v_g$ changes, $\delta{v_{\alpha\beta}}\equiv{v_\alpha-v_\beta}=\epsilon_\alpha-\epsilon_\beta$.
In the following we will usually take the particle-hole symmetric (phs) point given by $\epsilon_\alpha^\ast=-\frac{U_\alpha}{2}-\sum_{\beta\ne\alpha} U_{\alpha\beta}$ 
as the reference system. 

\subsection{Results for the double quantum dot}
\label{sub:ddot_results}

\begin{figure}
    \includegraphics[width=0.8\linewidth]{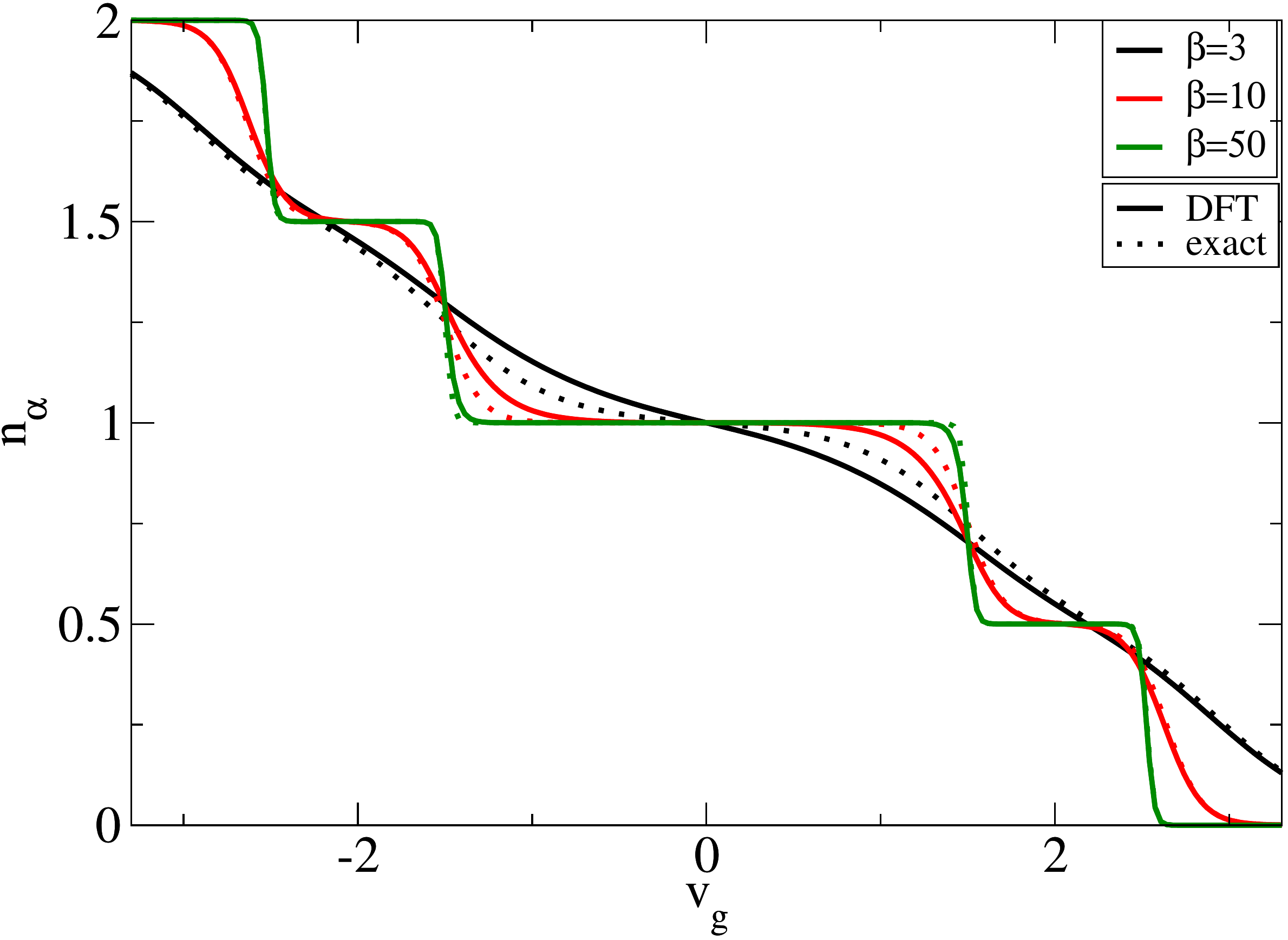} 
    \caption{Density ${\bm n}=(n_1,n_2)$ as function of the gate voltage $v_g$ for different temperatures when $U_{1}=U_{2}=3U_{12}>0$ (Regime I). 
      The DFT result (solid line) becomes on top of the GCE result (dashed line) in the low temperature regime. All energies in 
      units of the smallest interaction $U_{12}$.}
\label{fig:dqd_R1_equal}
\end{figure}

We now study the DQD, and start by considering the degenerate case in Regime I, 
i.e. $U_{1}=U_{2}>U_{12}$ where $\delta{N}=0$. Fig.~\ref{fig:dqd_R1_equal} compares the exact densities
with the ones computed in DFT using the Hxc potential for Regime I, Eq.~(\ref{eq:Hxc:I}), together with the parametrizations
of the SSM, Eq.~(\ref{eq:hxc_ssm}), and the CIM, Eq.~(\ref{eq:hxc_cim}), respectively. We see that the DFT results correctly describe all the features 
of the densities as a function of gate. At low temperatures, the width of the central step (around $v_g=0$) is given by $U_{i}$ while the other two step widths correspond 
to $U_{12}$. At higher temperatures our parametrization leads to moderate discrepancies in the slopes of the central step that disappear as the temperature approaches 
zero.

\begin{figure}
  \includegraphics[width=\linewidth]{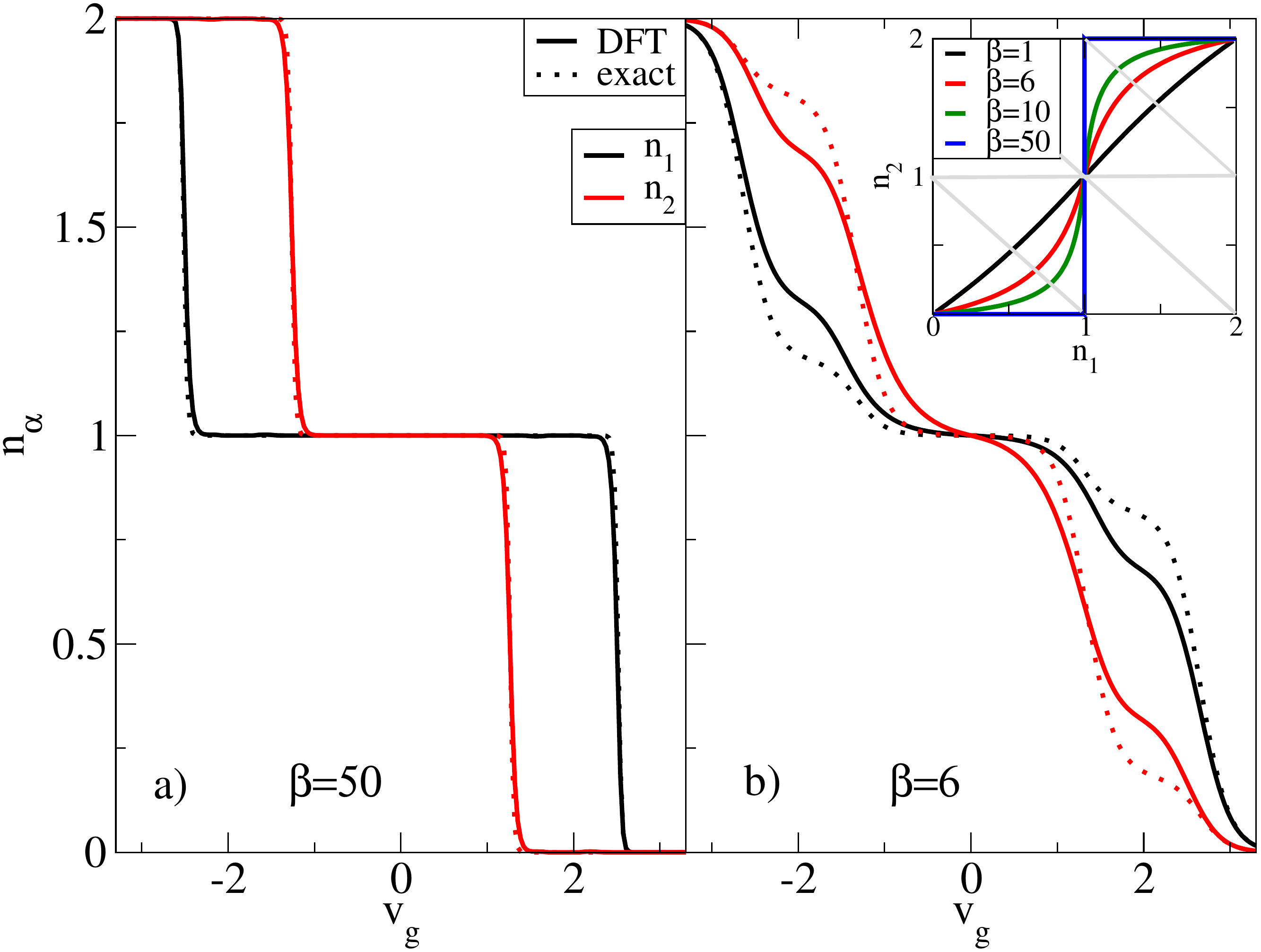} 
  \caption{
    Density${\bm n}=(n_1,n_2)$ as a function of the gate voltage $v_g$ for $U_{1}=3U_{12}$, $U_{2}=2.5U_{12}$ (Regime I) 
    for (a) low and (b) high temperatures. The inset of panel (b) shows the path in the $n_1-n_2$ plane as the gate is varied 
    for different temperatures. The grey lines show the steps of the CIM and SSM terms that appear in the Hxc potentials. 
    All energies in units of the smallest interaction $U_{12}$.
  }
\label{fig:dqd_R1}
\end{figure}

Next we consider the situation where the intra-orbital Coulomb repulsions are different, $U_{1}>U_{2}>U_{12}$. 
In Fig.~\ref{fig:dqd_R1}(a,b), the occupations $n_{i}$ are presented as a function of the gate $v_g$ for two 
different temperatures. At low temperatures [Fig.~\ref{fig:dqd_R1}(a)] at large negative gate voltage ($v_g<-2.5$) 
both orbitals of the DQD are completely filled ($n_\alpha\sim2$). As the gate is increased, first the orbital 
with the higher interaction ($U_1$) becomes half-filled around $v_g\sim-2.5$, and then around $v_g\sim-1.5$ also the orbital with the 
lower interaction ($U_2$) becomes half-filled. Upon further increase of the gate, the sequence of emptying is reversed, as
first the orbital with the higher interaction and thus lower gate ($v_2$) is emptied around $v_g\sim1.5$ and 
finally the orbital with lower interaction and thus higher gate ($v_1$) is emptied. At higher temperatures 
extra steps develop in the evolution of the density versus gate voltage, as can be seen in Fig.~\ref{fig:dqd_R1}(b).
The appearance of new steps can be understood by the path taken in the $n_1-n_2$ plane as the gate voltage changes,
shown in the inset of Fig.~\ref{fig:dqd_R1}(b) for different temperatures. At low temperatures the path essentially 
follows three straight line segments, along the lower border, across the plane and finally along the upper border,
thus avoiding extra steps of the CIM potential at $N=1$ and $N=3$. As the temperature increases the path becomes
smoother, and passes through the $N=1$ and $N=3$ steps of the CIM potential, leading to
the extra steps in the evolution of the densities at higher temperature. 
While for low temperatures the agreement of the DFT results with the exact ones is excellent, at higher temperatures 
deviations appear. Although DFT qualitatively captures the appearance of the extra steps in the evolution of the 
density versus gate voltage, their heights are not correctly reproduced in DFT. Presumably this discrepancy can be 
attributed to the development of a $\delta{N}$-dependence of the CIM potential at finite temperature, and will be addressed
in future work.

\begin{figure}
  \includegraphics[width=\linewidth]{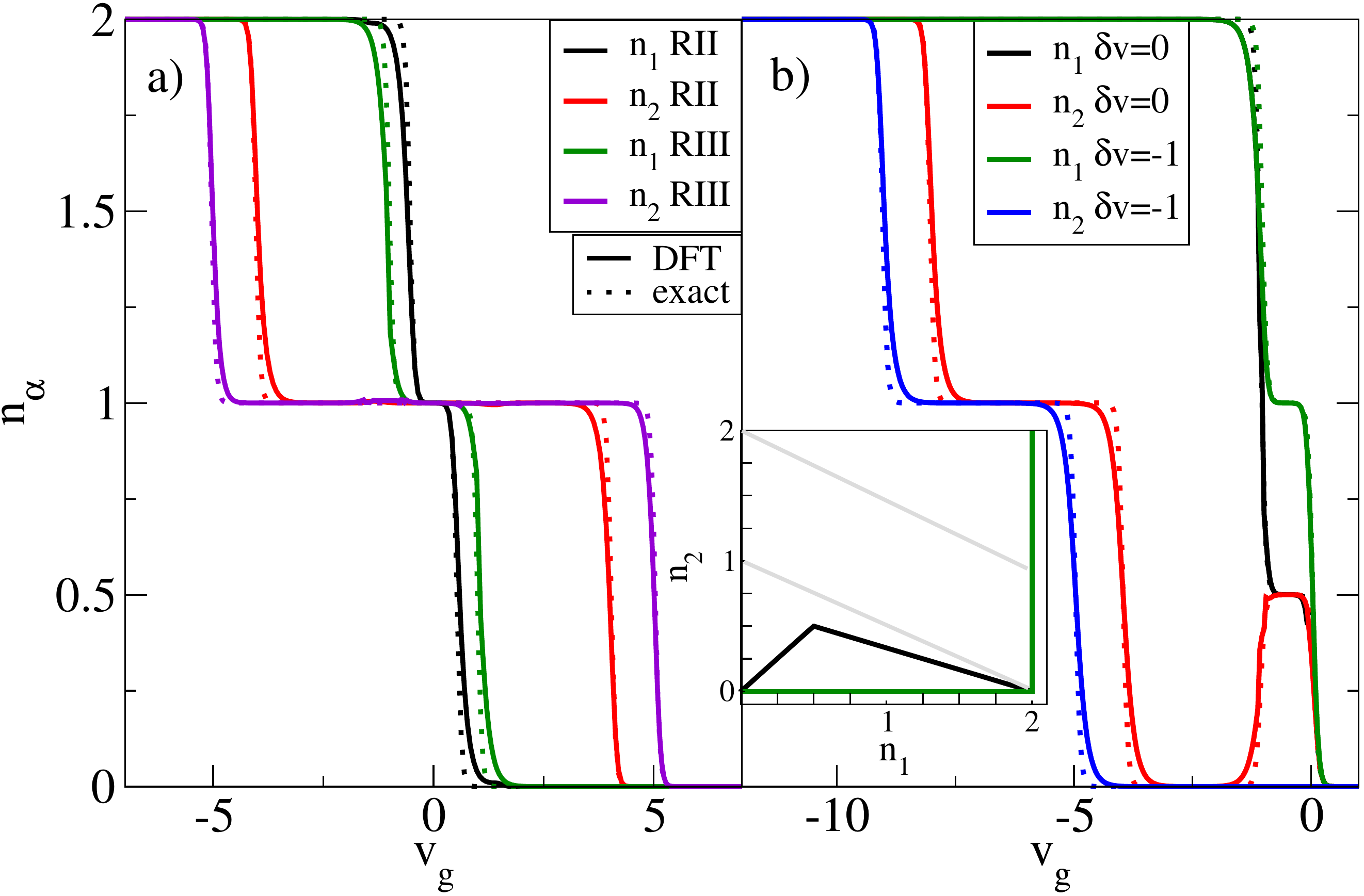} 
  \caption{
    (a) Comparsion of the evolution of the density ${\bm n}=(n_1,n_2)$ with the gate voltage $v_g$ 
    in Regime II ($U_2=2U_{12}=4U_1$) and Regime III ($U_2=U_{12}=4U_1$).
    (b) Comparison of density evolution for different two different values of the splitting $\delta{v}$ 
    in Regime II ($U_{12}=2U_1$). The inset shows the different paths in the density plane.    
    $\beta=20/U_1$ everywhere. All energies in units of $U_{1}$. 
  }
  \label{fig:R2-R3}
\end{figure}

Finally, we turn our attention to Regimes II and III, which are both characterized by the appearance of the peculiar
``Skew'' term in the Hxc potential. Fig.~\ref{fig:R2-R3}(a) directly compares the evolution of the density as a 
function of the gate in both regimes. As we can see the behaviour is actually quite similar for both regimes, and not 
so different from Regime I (cf. Fig.~\ref{fig:dqd_R1}): As the gate increases, first the orbital 
with the higher interaction (here $U_2$) becomes half-filled, and then the orbital with the 
lower interaction ($U_1$). Then upon further increase of the gate, the order of emptying is reversed.
Due to the higher inter-orbital interaction in Regime III, the width of the central plateau is increased for both orbitals.
Again, at low temperature the agreement with the exact results is excellent, but at higher temperatures moderate quantitative 
deviations occur (not shown).

In order to investigate the influence of the Skew term in the Hxc potential on the evolution of the densities,
we next concentrate on Regime II and explore different paths in the $n_1-n_2$ plane. To this end we fix the energy 
splitting $\delta{v}=v_1-v_2$ between the orbitals to different values while the total gate changes, i.e. $v_1 = \delta{v}+v_g$ 
and $v_2=v_g$. Fig.~\ref{fig:R2-R3}(b) shows the evolution of the density for two different values of $\delta{v}$ and 
correspondingly different paths in the $n_1-n_2$ plane (shown in the inset). 
For $\delta{v}=0$ we observe an interesting effect. As the gate increases, the occupation of orbital 2 decrease in 
two steps, first to half filled and then further to zero, while the first orbital remains fully occupied. 
Then around $v_{g}=-1$ the occupation of orbital 1 decreases abruptly to quarter filling, while now the 
occupation of orbital 2 increases again to quarter filling, $n_1=n_2=\sim0.5$. 
This non-monotonic behaviour of the occupation of orbital 2 is reminiscent of the so-called level occupation 
switching (LOS)\cite{Silvestrov:NJP:2007,Kleeorin:PRB:2017}.
We find similar behaviour in Regime III (not shown).

\subsection{Results for more than two orbitals}
\label{sub:mdot_results}

\begin{figure}
  \includegraphics[width=\linewidth]{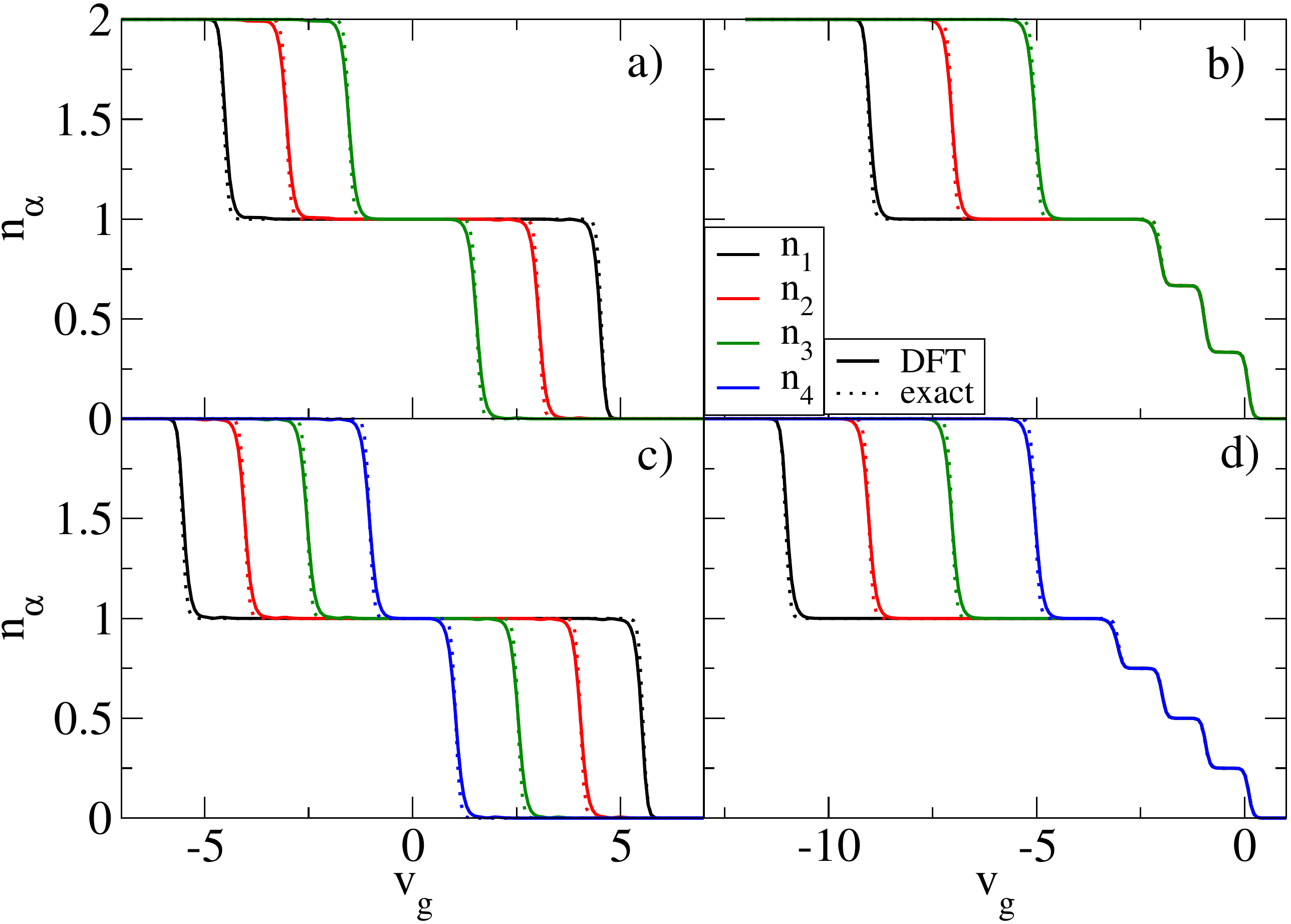} 
  \caption{Local occupations for the triple (a, b) and quadruple (c, d) QD as 
    function of the gate voltage. In the left panels the QD levels are taken at particle-hole 
    ($v_\alpha=\epsilon^\ast_\alpha+v_{g}$) and in the right the impurities level is set to zero ($v_\alpha=v_g$). 
    $\beta=20/U^\prime$ and $U_{1}=5U^\prime$, $U_{2}=4U^\prime$, $U_{3}=3U^\prime$, $U_{4}=2U^\prime$. All energies in units of $U^\prime$.}
\label{fig:3-4qd_R1}
\end{figure}

Finally we apply the generalization of the Hxc potential (\ref{eq:Hxc:multi-orbital}) for more than two orbitals to DFT calculations of multi-orbital QDs. 
Fig.~\ref{fig:3-4qd_R1} shows the evolution of the density ${\bm n}$ as a function of the applied gate voltage $v_g$ for three (a,b) and four-level (c,d) QD 
with all intra-orbital Coulomb repulsions $U_\alpha$ different and constant interdot repulsion $U^\prime $($U_1>U_2>\ldots>U^\prime$) at low temperature. 
In panels (a) and (c) the gate $v_g$ is applied w.r.t. the phs point,
i.e. $\epsilon_\alpha^\ast=-\frac{U_\alpha}{2}-\sum_{\beta\ne\alpha} U_{\alpha\beta}$. In this case the path in the three- or four-dimensional density space avoids the steps in the CIM Hxc potential away 
from half-filling ($N=\mathcal{M}$) resulting in only three plateaus in the density evolution with the gate, in a similar way as in the DQD [cf. Fig.~\ref{fig:dqd_R1}(a)].
On the other hand, in panels (b) and (d) (where $\epsilon_\alpha=0$ and thus $\delta{v}_{\alpha\beta}=0$) two (three) extra steps related to the inter-orbital Coulomb repulsions appear in the 
triple (quadruple) QD.
The agreement between the DFT and the exact results is remarkable in all cases, showing that the generalization of Eq.~(\ref{eq:Hxc:multi-orbital}) of the Hxc potential to more
than two orbitals is valid. Finding similar expressions for a more general choice of parameters will be the focus of future work.

\section{Conclusions}
\label{sec:conclusions}

In this work we have obtained Hxc potentials
for double quantum dots in the grand-canonical ensemble 
subject to generic density-density 
interactions and Hund's rule coupling by reverse-engineering 
from exact many-body solutions.
The structure of the Hxc potentials consists of ubiquitous steps
whose exact positions depend on the regime
defined by the interaction parameters. This structure can be 
understood and derived from an analysis of the stability diagrams. 
In a second step we were able to rationalize the step structure of
the Hxc potential by a decomposition of the interaction into 
basic components. This decomposition allows to write the Hxc potential 
of the system as a sum over basic Hxc potentials, which can be parametrized
in a straightforward manner. Importantly, the decomposition into basic potentials 
can be generalized to multi-orbital systems with more than two orbitals. 
DFT calculations employing the thus parametrized Hxc
potentials for double, triple and quadruple quantum dots show
excellent agreement with exact results at low temperatures.
At higher temperatures, we find moderate quantitative deviations from the exact results 
that we attribute to the modulation of step widths for finite $\delta{n}=n_1-n_2$
not captured in our parametrization.

The parametrization of the Hxc potential derived here could be directly applied e.g. to 
the description of single atoms where density-density and Hund's coupling are the 
dominating terms of the Coulomb interaction.
Possible further applications 
regard the description of transport through multi-orbital quantum dots or molecules coupled to leads. 
Due to the similarity between broadening by finite temperature on the one hand and 
finite coupling to the leads on the other hand, we expect that the Hxc potentials 
for finite coupling to the leads have a similar structure to the ones discussed here.\cite{KurthStefanucci:16}
One way to incorporate finite coupling to the leads in the Hxc potential is by 
introduction of an effective temperature.\cite{Sobrino:PRB:2019} 
For the description of non-equilibrium effects the i-DFT framework may be employed
which in addition to the Hxc gate potential requires a parametrization of the xc bias.\cite{StefanucciKurth:15}  
This would also allow one to compute many-body spectral functions of interacting multi-orbital systems.\cite{JacobKurth:18}

\begin{acknowledgments}
We acknowledge funding by the grant Grupos Consolidados UPV/EHU del Gobierno Vasco (IT1249-19) 
as well as the grant of the Ministerio de Econom\'ia, Industria y Competitividad, Gobierno de Espa\~na (MINECO) - Agencia Estatal de Investigaci\'on (FIS2016-79464-P) 
and European Regional Development Fund (FEDER), European Union.
\end{acknowledgments}

\bibliography{hxcfor2aim}

\end{document}